\documentclass[longauth]{aa}   
\def\linenumbers{}
\AtBeginDocument{\hoffset=12pt}

\usepackage{booktabs}
\usepackage{rotating}
\usepackage{xcolor}
\definecolor{revS}{rgb}{0,0.25,0.85}
\definecolor{revY}{rgb}{0,0.55,0.15}
\definecolor{revN}{rgb}{0.75,0,0.6}
\definecolor{tbdcol}{rgb}{0.9,0.45,0}
\definecolor{revT}{rgb}{0.85,0,0}
\definecolor{revR}{rgb}{0,0.5,0.55}
\definecolor{revP}{rgb}{0.55,0.3,0.05}
\newcommand{\rev}[1]{#1}
\newcommand{\revS}[1]{#1}
\newcommand{\revY}[1]{#1}
\newcommand{\revN}[1]{#1}
\newcommand{\revT}[1]{#1}
\newcommand{\revR}[1]{#1}
\newcommand{\revP}[1]{#1}
\definecolor{revV}{rgb}{0.35,0.45,0.0}
\newcommand{\revV}[1]{#1}
\definecolor{revF}{rgb}{0.8,0.4,0.0}
\newcommand{\revF}[1]{#1}

\providecommand{\pss}{Planet.~Space~Sci.}
\providecommand{\psj}{Planet.~Sci.~J.}
\providecommand{\areps}{Annu.~Rev.~Earth Planet.~Sci.}
\providecommand{\science}{Science}
\providecommand{\natas}{Nat.~Astron.}

\begin{document}

\title{Centimetre-sized meteoroids from the active asteroid (3200) Phaethon: lunar impact flashes during the 2018 and 2025 Geminids}

\author{S.~Abe\inst{1,2}
   \and N.~Arima\inst{1}
   \and M.~Nishikata\inst{1}
   \and Y.~Araki\inst{1}
   \and T.~Aoki\inst{1}
   \and S.~Akutsu\inst{1}
   \and J.~Okuyama\inst{1}
   \and R.~Fuse\inst{3}
   \and D.~Fujii\inst{4}
   \and S.~Sato\inst{5}
   \and T.~Tanaka\inst{6}
   \and R.~Yamada\inst{7}
   \and S.~Tanaka\inst{8}
   \and H.~Shiraishi\inst{8}
   \and K.~Onodera\inst{9}
   \and T.~Kawamura\inst{10}
   \and M.~Yanagisawa\inst{11}}

\institute{Department of Aerospace Engineering, Nihon University, 7-24-1 Narashinodai,
           Funabashi, Chiba 274-8501, Japan; \email{abe.shinsuke@nihon-u.ac.jp}
     \and  Space Science Research Unit (NU-SX), Research Institute of Science and Technology,
           Nihon University, 1-8-14 Kanda Surugadai, Chiyoda-ku, Tokyo 101-8308, Japan
     \and  Department of Aeronautics and Astronautics, The University of Tokyo, 7-3-1 Hongo,
           Bunkyo-ku, Tokyo 113-8656, Japan
     \and  Hiratsuka City Museum, 3-9-23 \revF{Sengen-cho}, Hiratsuka, Kanagawa 254-0041, Japan
     \and  Department of Oceanic Architecture and Engineering, Nihon University,
           7-24-1 Narashinodai, Funabashi, Chiba 274-8501, Japan
     \and  Institute of Quantum Science, Nihon University,
           7-24-1 Narashinodai, Funabashi, Chiba 274-8501, Japan
     \and  The University of Aizu, Tsuruga, Ikki-machi, Aizu-Wakamatsu, Fukushima 965-8580, Japan
     \and  Institute of Space and Astronautical Science, Japan Aerospace Exploration Agency,
           3-1-1 Yoshinodai, Chuo-ku, Sagamihara, Kanagawa 252-5210, Japan
     \and  \rev{Institute for Planetary Materials, Okayama University, 827 Yamada, Misasa,
           Tohaku-gun, Tottori, 682-0193, Japan}
     \and  Universit\'e Paris Cit\'e, Institut de physique du globe de Paris, CNRS,
           1 rue Jussieu, 75005 Paris, France
     \and  The University of Electro-Communications, 1-5-1 Chofugaoka, Chofu,
           Tokyo 182-8585, Japan}

\date{Received XXX; accepted XXX}

\authorrunning{S.~Abe et al.}
\titlerunning{Centimetre-sized Phaethon meteoroids: lunar impact flashes in the 2018 and 2025 Geminids}

\abstract
{Centimetre-sized meteoroids released by the active asteroid (3200) Phaethon, parent of the Geminid meteor shower, probe the ejection of refractory material from a comet--asteroid transition object and the age of the stream; lunar impact flashes (LIFs) are the only ground-based technique that samples this size range.}
{We determine the impact flux, mass and size distributions, and expected crater sizes of Geminid meteoroids on the Moon from campaigns at the 2018 and 2025 maxima, and test the stability of the centimetre-sized population over seven years.}
{Both campaigns used the Nihon University 0.40~m ``GUNDAM'' telescope with CMOS video cameras at 60~fps, with candidates validated at independent stations. A common aperture-photometry pipeline calibrated on standard stars in the Cousins $R$ band gave peak magnitudes to 0.1--0.25~mag, from which energies, masses, crater sizes, and shower-association probabilities follow.}
{Eleven LIFs were detected in \revN{3.3}~h in 2018 and nine in 2.2~h in 2025, all with Geminid probabilities of \revP{0.87}--0.96 (\revP{0.3}--0.7 with a mass-based prescription). With \revP{monitored areas measured on the video frames}, the campaigns give consistent fluxes of \revN{$4.2\times10^{-7}$} and \revP{$5.2\times10^{-7}$}$\,\mathrm{km^{-2}\,h^{-1}}$ (merged: \revN{$4.6\times10^{-7}$}$\,\mathrm{km^{-2}\,h^{-1}}$, mass flux $2.8\times10^{-5}$~g\,km$^{-2}$\,h$^{-1}$). For a luminous efficiency of $1.4\times10^{-3}$ the impactors have masses of \revN{10--320}~g, diameters of \revN{1.8--5.9}~cm, and predicted crater rim diameters of 2--5~m. The combined sample gives a maximum-likelihood mass index $s=1.76\pm0.17$, size indices $\beta_{1}=3.3\pm0.5$ and $\beta_{2}=3.6\pm0.6$, and a population index $r=2.0\pm0.3$, markedly shallower than the NELIOTA all-flash values; a single power law with this slope connects the Global Meteor Network millimetre flux to the LIF centimetre flux.}
{The Geminid stream delivers an excess of large particles relative to the sporadic background, consistent with an asteroidal origin, and its stable centimetre-sized flux over seven years constrains stream-formation models with ages of $2\times10^{3}$--$2\times10^{4}$~yr. Identifying a predicted metre-class crater in LRO temporal imaging would calibrate the luminous efficiency at a known velocity.}


\maketitle

\section{Introduction}

Meteoroids in the size range from about 30~$\mu$m to 1~m originate from
comets and asteroids. When a millimetre-sized meteoroid enters the
terrestrial atmosphere at several tens of \mbox{km\,s$^{-1}$} it produces a
visible meteor, whereas a meteoroid larger than about a centimetre that
strikes the lunar surface produces a short ($0.01$--$0.1$~s) optical
transient known as a lunar impact flash (LIF). Typical LIFs radiate from
visible to near-infrared wavelengths and are observed as point sources of
5--11~mag on the night side of the lunar disc
\citep{Ortiz1999,Madiedo2019book}. Because a telescope monitoring the
non-sunlit hemisphere of a half Moon surveys an effective collecting area
roughly two orders of magnitude larger than a single ground-based fireball
station \citep{Plane2012}, LIF monitoring efficiently samples the
poorly constrained centimetre-to-decimetre size range of the meteoroid
population entering cislunar space. Systematic surveys, notably the NASA
Marshall Space Flight Center programme \citep{Suggs2014} and the
ESA-supported NELIOTA project at Kryoneri Observatory
\citep{Xilouris2018,Liakos2020,Liakos2024}, have accumulated several
hundred validated flashes, and recent developments extend the technique to
daytime short-wave-infrared observations \citep{Sheward2024}, to
high-speed ($>$100~fps) photometry \citep{Giancono2026}, and to the
interpretation of light-curve morphology in terms of mare versus highland
target properties \citep{Athanasopoulos2026}. Spectroscopic observations
constrain blackbody temperatures of 2000--4000~K and, in the early flash
phase, an additional $\sim$6000~K vapour component
\citep{Yanagisawa2021,Yanagisawa2025}, while the meteoroid composition can
also be probed through the emission \citep{Avdellidou2019}. Continuous LIF
monitoring further supports risk assessment for future human and robotic
lunar-surface activities, and impact events with known source locations
and times provide seismic calibration signals: \rev{about 6\% of the more than
33\,000 seismic events recorded by the Apollo seismometers} are attributed to
meteoroid impacts \rev{\citep{Nakamura1981,Onodera2024}}, and a single travel-time datum from a located impact
usefully constrains lunar interior models \citep{Yamada2011}. Long-duration
monitoring from cislunar orbits, such as a near-rectilinear halo orbit
around the Earth--Moon L$_{2}$ point, has been demonstrated conceptually by
the EQUULEUS/DELPHINUS mission \citep{Funase2020}.

The Geminid meteor shower is at present the strongest annual shower, with
a zenithal hourly rate (ZHR) reaching $\sim$150. Its parent body is the
B-type active asteroid (3200) Phaethon \citep{Jewitt2010}, a
comet--asteroid transition object on a peculiar orbit with eccentricity
$e=0.89$, perihelion distance $q=0.14$~au, and period $P=1.44$~yr. The
formation and evolution of the Geminid stream, and the relation of
Phaethon to its likely split companion 2005~UD, constitute the
``Phaethon--Geminid stream complex'' problem
\citep{Ohtsuka2006,Devogele2020,MacLennan2024}. Dynamical models of the
stream imply formation ages ranging from $\sim$2000~yr up to
$\sim$18\,000~yr if the particles were released by rotational instability
\citep{Ryabova2012,Cukier2023,Jo2024}, and white-light imaging by the
Parker Solar Probe WISPR instrument has revealed the dense core of the
Geminid stream near Phaethon's orbit, providing a new observational
benchmark for these models \citep{Battams2025}. Near perihelion the
surface temperature of Phaethon exceeds 1000~K, and the depletion of
sodium in Geminid meteoroids has been demonstrated spectroscopically
\citep{Abe2020}. The OSIRIS-REx mission unexpectedly observed repeated
ejections of centimetre-sized pebbles from the B-type near-Earth asteroid
(101955) Bennu \citep{Lauretta2019}, showing that centimetre-scale mass
loss is a generic feature of active asteroids. Phaethon is the flyby
target of the JAXA DESTINY$^{+}$ mission, \revS{whose Dust Analyser will
measure the flux, dynamics, and composition of sub-micrometre- to
micrometre-sized dust in situ} (\citealt{Ozaki2022,Arai2024}\revS{; \citealt{Simolka2024}}); LIF
observations of the Geminids probe the complementary centimetre size
range of the same parent body.

In this paper we report the complete analysis of LIF observations
obtained at Nihon University during the maxima of the 2018 and 2025
Geminid showers. The 2018 campaign, conducted jointly with the University
of Electro-Communications (UEC), Aizu University, ISAS/JAXA, and Lulin
Observatory (Taiwan), yielded eleven confirmed flashes whose low-dispersion
spectra were published by \citet{Yanagisawa2021}; the \revR{analysis of the
direct images (0.8\arcsec\ per pixel)} summarised here was first reported by \citet{Abe2024}.
The 2025 campaign yielded nine flashes, all consistent with a Geminid
origin. We derive impact fluxes, mass and size distributions, and expected
crater diameters with a common methodology, and discuss the implications
for the centimetre-sized particle population of the Phaethon--Geminid
stream complex.

\section{Observations}

\subsection{Instrumentation}

Both campaigns used the dedicated LIF telescope system (nicknamed the
``GUNDAM'' telescope) on the Funabashi campus of the College of Science
and Technology, Nihon University
($140\degr03\arcmin32\arcsec$E, $+35\degr43\arcmin31\arcsec$N, 40~m
altitude). The 0.40~m Officina Stellare RiFast~400 Cassegrain telescope ($f/3.8$;
a co-mounted 0.20~m wide-field telescope was not used in this analysis)
is carried on a Showa Kikai 35EF fork equatorial mount and tracks the Moon
(Table~\ref{tab:instruments}). Cooled, high-sensitivity CMOS cameras
(quantum efficiency $>$20\% over 350--800~nm) were attached at the
Cassegrain foci without filters. In 2025 the camera was synchronised
to GPS time; in 2018 the frames were time-stamped by the PC clock\revY{ of
the recording computer}\revV{, which was synchronised to an NTP server}\revY{. The PC clocks at UEC were set manually, and the
times of the simultaneously observed flashes listed by
\citet{Yanagisawa2021} are those of the Funabashi data; the coincidence
of the detections at the two sites was established to within $\pm1$~s}. Video
was recorded in SER format at \revV{a nominal} 60 frames~s$^{-1}$ \revV{(recorded rates: 58.1--61.0 in 2018, 60.2 in 2025)} with an exposure time of
16.67~ms per frame, and impact-flash candidates were extracted after the
observations with \revV{software developed in-house for both campaigns}.

\begin{table}
\caption{GUNDAM telescope and camera configuration.}
\label{tab:instruments}
\centering
\begin{tabular}{lc}
\toprule
Telescope            & Officina Stellare RiFast~400 \\
Aperture (mm)        & 400 \\
Focal length (mm)    & 1520 \\
Focal ratio          & 3.8 \\
Pixel resolution (\arcsec) & 0.8 \\
Field of view (\arcmin)  & $25.6\times16.1$ \\
\midrule
Camera (2018)        & ZWO ASI174MM-Cool \\
Camera (2025)        & QHY174GPS-M \\
Sensor format (pix)  & $1936\times1216$ ($5.86\,\mu$m) \\
Exposure / frame rate & 16.67 ms / \revV{$\simeq$}60 fps \\
Digitisation         & 8 bit (2018) / 16 bit (2025) \\
\bottomrule
\end{tabular}
\tablefoot{Pixel resolution and field of view refer to the camera format
given in the lower part of the table. The QHY174GPS-M was operated with a
gain of 0.0672~e$^{-}$/ADU and a readout noise of 2.18~e$^{-}$.}
\end{table}

\begin{table*}
\caption{Circumstances of the two Geminid campaigns.}
\label{tab:circumstances}
\centering
\footnotesize
\setlength{\tabcolsep}{3.5pt}
\begin{tabular*}{\textwidth}{@{\extracolsep{\fill}}lcc}
\toprule
 & 2018 & 2025 \\
\midrule
Date (UT)                  & 15 Dec 2018      & 14/15 Dec 2025 \\
Observing window (UT)      & \revY{08:20}--11:40     & 18:50--21:00 \\
Effective duration (h)     & \revN{3.3}        & 2.2 \\
Lunar age$^{a}$ (d)              & \revY{8.1} (waxing)     & \revN{24.5} (waning) \\
Illuminated fraction$^{a}$       & 0.49             & \revN{0.23} \\
Earth--Moon distance$^{a}$ (km) & $3.956\times10^{5}$ & $4.005\times10^{5}$ \\
Geminid ZHR$_{\max}$       & $155.3\pm10.5$   & $142.8\pm8.7$ \\
Confirmed flashes          & 11               & 9 \\
Confirming station(s)      & UEC (Chofu)      & Hiratsuka, Asakuchi, Seimei \\
Photometric standard star   & HD~221835        & HD~117544 \\
Extinction $k$ (mag airmass$^{-1}$) & 0.20 (assumed)  & 0.28 (measured) \\
Peak magnitude range ($R$)  & 5.9--9.5         & 7.6--9.7 \\
Monitored night-side area$^{b}$ ($10^{6}$~km$^{2}$) & \revP{$7.8\pm0.3$} & $8.1\pm0.3$ \\
Fraction of the visible night side & 0.82 & 0.62 \\
Exposure $A\,T$ ($10^{6}$~km$^{2}$\,h) & \revN{26.1} & 17.5 \\
Impact flux (km$^{-2}$\,h$^{-1}$) & \revN{$4.2\times10^{-7}$} & \revP{$5.2\times10^{-7}$} \\
\bottomrule
\end{tabular*}
\tablefoot{$^{(a)}$ At mid-observation\revY{; the lunar age is counted from the preceding new moon (8.0--8.2~d and 24.5--24.6~d over the two runs)}; the Earth--Moon distance is topocentric. $^{(b)}$ True
surface area of the Earth-facing night side inside the 0.40~m field of
view, time-weighted over the run (mean $\pm$ standard deviation over the
clips; Sect.~\ref{sec:area}). Maximum ZHR values are from the International Meteor
Organization visual-observation analyses \citep[2018:][2025: IMO Geminids
live campaign]{Molau2020}.}
\end{table*}

\subsection{The 2018 campaign}
\label{sec:obs2018}

The 2018 observations were performed on 15 December 2018 from \revY{17:20} to
20:40 JST (\revY{08:20}--11:40~UT), about half a day after the visual maximum of
the shower (ZHR $=155.3\pm10.5$ at solar longitude
$\lambda_{\sun}=262.358\degr$; \citealt{Molau2020}). The Moon was a
\revY{8.1-day-old (8.0--8.2~d during the run)} waxing half Moon, with the Geminid radiant impacting the
Earth-facing night side -- nearly ideal geometry. Eleven flashes
(designated \revY{B--E and G--M} following \citealt{Yanagisawa2021}) were detected in
\revN{3.3}~h. All eleven events were simultaneously recorded at UEC
(Chofu, Tokyo), $\sim$50~km from Funabashi, excluding satellite glints
and electronic noise. \revP{The selenographic coordinates of the flashes were re-measured in this
work by registering the earthshine-lit lunar surface in each video clip
on the LROC Wide Angle Camera mosaic (Sect.~\ref{sec:area}),
with uncertainties of $0.2\degr$ ($0.5\degr$ and $1.9\degr$ in longitude for
flashes L and H, which lie near and on the limb); they agree with the earlier
positions of \citet{Abe2024} to within 10--23~km (great-circle separations)
for seven flashes, the exceptions being K and L near the western limb
(61 and 66~km), M (65~km), and H (182~km, within its earlier $\pm5\degr$
uncertainty in longitude).}
\revV{The new positions are roughly ten times more precise than the preliminary positions of}
\citet{Yanagisawa2021}, which were referenced to the earthshine-lit
features Grimaldi and Aristarchus.
The distribution of the flashes on the lunar disc is shown in Fig.~\ref{fig:positions}a.

\subsection{The 2025 campaign}
\label{sec:obs2025}

\begin{figure*}
\centering
\includegraphics[width=0.72\textwidth]{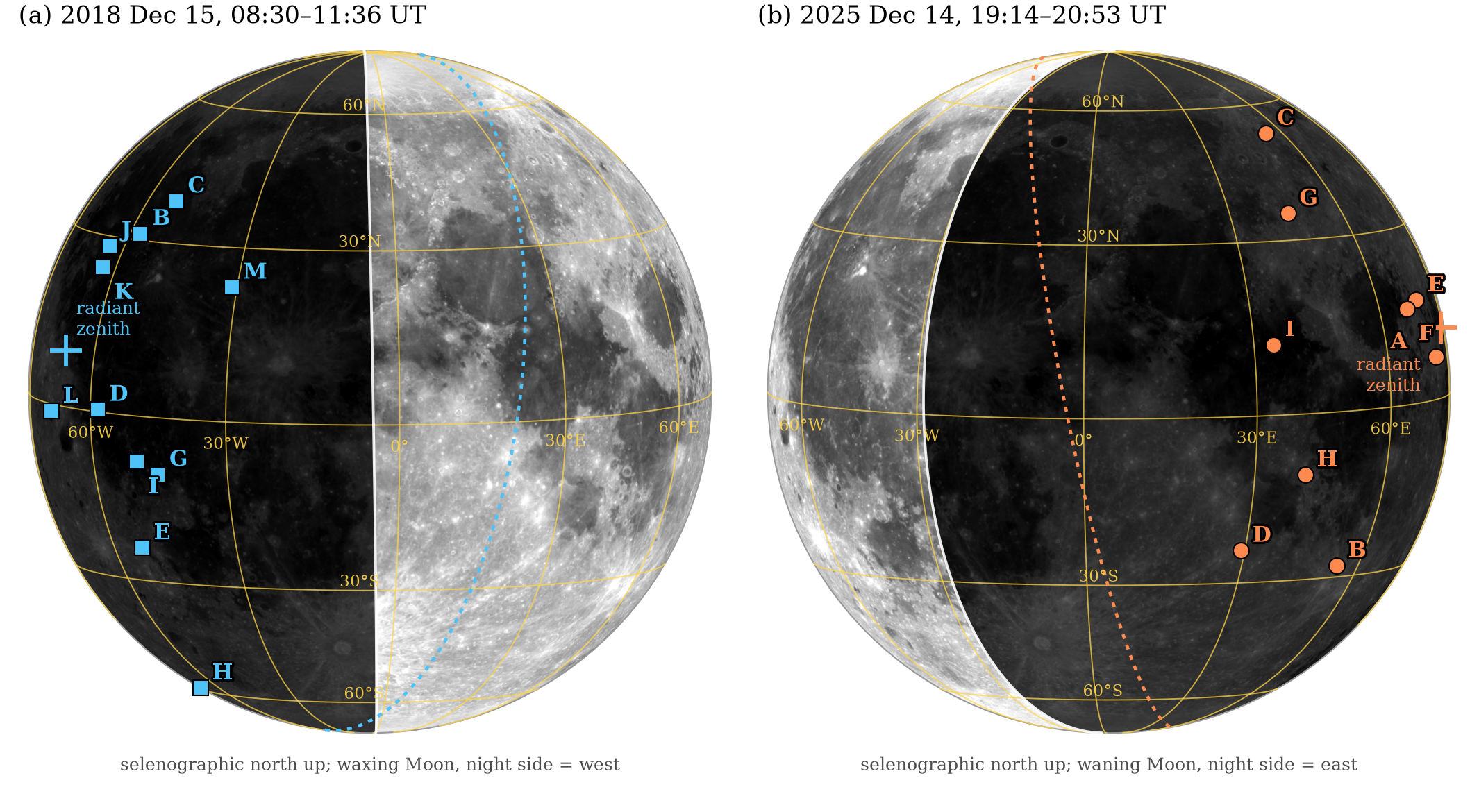}
\caption{Impact locations of the Geminid lunar impact flashes on the
LROC Wide Angle Camera mosaic (orthographic projection centred on the
sub-Earth point at mid-campaign; selenographic north up). (a) Eleven
flashes \revY{B--E and G--M} of 15 December 2018. (b) Nine flashes A--I of
14 December 2025. The night side is shaded and the white line is the
terminator. The cross marks the sub-radiant point of the Geminids (where
the radiant is at the zenith), and the dotted great circle is the locus
where the radiant is on the local horizon. The coordinates and their
uncertainties are listed in Table~\ref{tab:events}.}
\label{fig:positions}
\end{figure*}

The 2025 observations were performed in the early morning of 15 December
2025 JST (03:50--06:00 JST, i.e. 14 December 18:50--21:00~UT), with a
\revN{24.5}-day-old waning Moon. Nine flashes (A--I) were detected in the
2.17~h of lunar recording, all within 19:14--20:53~UT.
Simultaneous video was obtained at Hiratsuka (Kanagawa; 3~ms and 10~ms
exposures) and Asakuchi (Okayama); the confirmation status of each flash
in our own reduction of the Hiratsuka data is given in the notes to
Table~\ref{tab:events}. Flash E, a marginal detection at Funabashi,
was also recorded simultaneously with the 3.8~m Seimei telescope of
Kyoto University at Okayama Observatory \citep{Arima2026}. The maximum ZHR of the 2025 Geminids was
$142.8\pm8.7$; the ZHR at the epoch of each flash was estimated from the standard post-maximum decline of the activity profile (Sect.~\ref{sec:assoc}). Flash I, recorded by all three cameras, is shown in
Fig.~\ref{fig:multicam}, and the impact locations are shown in
Fig.~\ref{fig:positions}b. \revP{The coordinates were re-measured in this work
by the same registration on the LROC mosaic (Sect.~\ref{sec:area}); they agree with the chart-based positions
of five flashes (A, C, D, H, and I) to within 11--102~km but
correct those of flashes B (by 187~km), G (by 331~km), and F (by 1900~km);
the earlier values of F and G had latitude and longitude interchanged.
For flash E we adopt the position measured on the simultaneous
Seimei/TriCCS image \citep{Arima2026}, so that both papers refer to the
same coordinates; our independent registration of the Funabashi clip
($17.5\degr$N, $74.4\degr$E) agrees with it to within $0.3\degr$ (9~km).}

\section{Data reduction and analysis}

\subsection{Photometry and calibration}
\label{sec:phot}

Both data sets were re-reduced with a single aperture-photometry pipeline
written for this work. After dark subtraction and flat-fielding (2025), a
median frame of each clip was used as the reference image and flashes were
located with a matched filter on the residual cube\revR{ (each residual frame,
frame minus median, was smoothed with a Gaussian kernel of
$\sigma=1.2$~pixels matched to the flash image and divided by the
per-pixel temporal noise, the median absolute deviation over the clip)}; candidates coinciding
with saturated sunlit terrain or lacking a transient time profile were
rejected. Each flash was measured in every frame with a circular aperture
of 9~pixels ($7.2\arcsec$) radius and a sigma-clipped background annulus of
16--24~pixels, re-centred on the flash centroid; a constant baseline
(the median flux outside the event) was subtracted to remove the gradient
of the earthshine background. The two faintest 2018 flashes (D and J) were
re-measured with a 5-pixel aperture and corrected to the 9-pixel scale
with a growth curve measured on five bright, unsaturated flashes
($1.13\pm0.09$). Frames in which the 8-bit 2018 camera saturated (flashes
B, H, and L) were recovered by fitting a two-dimensional Gaussian to the
unsaturated wings of the image. The $R$-band magnitude of a flash is
\begin{equation}
R = \mathrm{ZP} - 2.5\log_{10} f' - kX ,
\label{eq:pogson}
\end{equation}
where $f'$ is the baseline- and background-subtracted flux per frame,
$X=\sec z$ is the airmass at zenith angle $z$, $k$ the atmospheric
extinction coefficient, and ZP the photometric zero point determined from
standard stars observed with the same camera settings. The unfiltered
CMOS passband (350--800~nm) is close to a broad $R$ band for the
2000--4000~K thermal spectra of the flashes, and the $R$ band was chosen
for direct comparison with previous surveys \citep{Suggs2014,Liakos2020};
no colour term was applied.

In 2018 the G8\,III star HD~221835 ($R_{C}=5.94\pm0.04$, from Gaia~DR3
photometry with the transformation of \citealt{Riello2021}) was recorded
in two clips at $X=1.38$--$1.42$ with the same gain and exposure, giving
$\mathrm{ZP}=15.99\pm0.08$ for $k=0.20\pm0.08$~mag per airmass; the Moon
was observed at $X=1.38$--$1.94$, so the uncertainty of $k$ contributes
less than 0.05~mag. In 2025 the F0\,IV/V star HD~117544
($R_{C}=6.71\pm0.05$) remained within $1\degr$ of the Moon throughout the
run and is present in the video cubes of flashes B and C as well as in
three dedicated clips; five measurements at $X=2.00$--$2.46$ give
$\mathrm{ZP}=23.48\pm0.06$ per 16.67~ms frame with $k=0.2822$ measured
during the same night (a free fit to the five points gives
$k=0.18\pm0.15$, consistent with the adopted value). The total
uncertainties of the peak magnitudes are 0.08--0.25~mag, dominated by
photon noise for the faint flashes and by the zero point for the bright
ones. Two independent checks support the calibration: \revY{the eight 2018
flashes recorded with the UEC 0.45~m telescope agree with the $R$
magnitudes measured there simultaneously (Table~2 of
\citealt{Yanagisawa2021}) with a mean difference (this work minus UEC)
of $+0.38$~mag and an rms scatter of 0.30~mag (Fig.~\ref{fig:uec}); the
three flashes recorded at UEC only with the 0.28~m telescope (E, H, M)
appear fainter there by up to 2~mag, because their spectra were truncated
at 750~nm to avoid second-order contamination, which biases the
spectrally derived $R$ magnitudes towards faint values
\citep{Yanagisawa2021}.} The 2025 flashes recorded simultaneously with a
3~ms-exposure camera at Hiratsuka, calibrated on the same star, agree
with the Funabashi magnitudes (16.67~ms equivalent) with a mean difference
of $-0.2$~mag and a scatter of 0.4~mag (Appendix~\ref{app:phot}). The 2018 peak magnitudes are on
average 0.7~mag brighter (rms 0.5~mag) than the preliminary values of
\citet{Abe2024}, which were tied to the Gaia~$G$ band through a
colour-independent offset; the present values supersede them.

\revV{The observing-site conditions (extinction, scintillation, and image
size) are summarised in Appendix~\ref{app:phot}. The faint-end cutoff of
the detected flashes, $R_{\lim}\simeq9.5$--9.7, was essentially identical
in the two campaigns.} \revR{The site is therefore adequate for LIF photometry, its
sensitivity being limited mainly by the earthshine background and by
scintillation rather than by seeing. The faint-end cutoff of the
detected flashes, $R_{\lim}\simeq9.5$--9.7 (single-frame $3\sigma$
limit $R\simeq9.8$--10.2), was essentially identical in the two
campaigns.}

\subsection{Monitored lunar area}
\label{sec:area}

The field of view of the 0.40~m telescope ($25.6\arcmin\times16.1\arcmin$)
covers only part of the lunar disc, and the telescope was re-pointed and
drifted during each run, so the monitored area was measured directly on
the video. \revP{For every flash clip the position of the lunar disc in the frame and
the orientation of the camera on the sky were determined by registering
the earthshine-lit lunar surface on the LROC Wide Angle Camera albedo
mosaic. The solution uses neither the pixel position of the flash nor
any earlier coordinates, and its accuracy is about one pixel
(Appendix~\ref{app:reg}).}
With the lunar rotation model of Sect.~\ref{sec:assoc} the
frame was then projected on a $0.5\degr$ grid of the lunar surface and the
true surface area of the Earth-facing night side inside the frame was
integrated for each clip. The result,
$A=7.8\pm\revP{0.3}\times10^{6}$~km$^{2}$ in 2018
(range 7.3--8.1$\times10^{6}$~km$^{2}$,
i.e. 82\% of the visible night side) and
$A=8.1\pm0.3\times10^{6}$~km$^{2}$ in 2025
(range 7.5--8.5$\times10^{6}$~km$^{2}$,
62\% of the much larger night side of the waning Moon),
is the time-weighted mean over the run, each clip representing the
interval to the mid-points of its neighbours. The corresponding
collecting areas projected on the Geminid radiant,
$A_{\perp}=\sum\cos\varphi\,\mathrm{d}A$, are
$5.5\times10^{6}$ and \revP{$4.8\times10^{6}$}~km$^{2}$\revR{, where the sum runs over
the surface elements $\mathrm{d}A$ ($0.5\degr$ grid cells) of the monitored
area and $\varphi$ is the angle between the local surface normal and the
direction of the Geminid radiant (corrected for the orbital velocity of
the Moon), the incidence angle of Table~\ref{tab:events}}.
\revP{The same solution converts the pixel position of each flash (the centroid
of the peak frame minus the median frame) directly
into selenographic coordinates (Table~\ref{tab:events}). Their
uncertainties combine the registration residuals, the scatter of the
rotation angle, the plate scale, and the flash centroid, with a floor of
$0.2\degr$ (6~km); they are larger in longitude for the flashes near the
limb. These coordinates supersede the earlier chart-based values
(Sects.~\ref{sec:obs2018} and \ref{sec:obs2025}). As an independent check,
the registration of the Funabashi clip of flash 2025~E reproduces to within
$0.3\degr$ the position measured on the simultaneous Seimei/TriCCS image
\citep{Arima2026}.} The uncertainty of the monitored area
is dominated by the pointing variations between the sampled epochs ($\pm$3\% in 2018, $\pm$4\% in
2025) and is smaller than the Poisson uncertainty of the flash counts.

\subsection{Impact energy, mass, and size}
\label{sec:energy}

The luminous energy radiated in the $R$ band during one video frame is
\begin{equation}
E_{\mathrm{lum}} = f_{\lambda}\,\Delta\lambda\, f \pi d^{2}\, t ,
\label{eq:elum}
\end{equation}
with
\begin{equation}
f_{\lambda} = 10^{-7} \times
10^{-\left(\frac{R + 21.1 + \mathrm{ZP}_R}{2.5}\right)}
\quad [\mathrm{J\,cm^{-2}\,s^{-1}}\,\revS{\mbox{\AA}^{-1}}],
\label{eq:flambda}
\end{equation}
where $\Delta\lambda = 1607$~\AA\ is the $R$-band width,
$\mathrm{ZP}_R=0.555$ its zero point \citep{Bessell1998}, $d$ the
topocentric Earth--Moon distance from Funabashi at the time of each flash
($3.956$--$3.965\times10^{10}$~cm in 2018; $4.000$--$4.012\times10^{10}$~cm
in 2025), and
$t=16.67$~ms the frame exposure time, which approximates the flash
duration. The factor $f=2$ assumes isotropic radiation of the impact
plume into the $2\pi$~sr hemisphere above the surface. The impactor
kinetic energy follows from the luminous efficiency $\eta$,
\begin{equation}
E_{k} = E_{\mathrm{lum}}/\eta ,
\label{eq:ek}
\end{equation}
for which \citet{Swift2011} derived from hypervelocity-impact
experiments and lunar monitoring the velocity-dependent form
\begin{equation}
\eta = 1.5\times10^{-3}
\exp\!\left\{-\frac{(9.3~\mathrm{km\,s^{-1}})^{2}}{v^{2}}\right\}
\label{eq:eta}
\end{equation}
later adopted by \citet{Suggs2014} \citep[see also][]{BellotRubio2000}. The luminous efficiency
remains the dominant source of systematic uncertainty: laboratory impacts
into quartz sand at 0.01~Pa give $\eta\simeq3\times10^{-4}$ at
6.5~km\,s$^{-1}$ \citep{Fuse2020}, Geminid LIF analyses give
$\eta\sim10^{-3}$ \citep{Ortiz2015}, and values as high as
$6\times10^{-3}$ have recently been proposed from flash--crater matching
(e.g. \revR{\citealt{Sheward2025}}). Following \citet{Abe2024} we adopt the
nominal value $\eta=1.4\times10^{-3}$ from Eq.~(\ref{eq:eta}) with the
Geminid lunar impact velocity \revY{$v=33.9$~km\,s$^{-1}$. This value
follows from the geocentric velocity $V_{g}=33.8$~km\,s$^{-1}$ of
video-observed Geminids \citep{Jenniskens2016} by adding in quadrature
the lunar escape speed (2.4~km\,s$^{-1}$) and the Earth's gravitational
potential at the lunar distance; the frequently quoted value
$V_{\infty}=35$~km\,s$^{-1}$ is the speed at the top of the Earth's
atmosphere and includes the acceleration by the Earth's gravity. The
orbital motion of the Moon ($\sim$1~km\,s$^{-1}$) changes the impact
speed to 34.8~km\,s$^{-1}$ in 2018 and 32.9~km\,s$^{-1}$ in 2025; we
adopt the mean value for both campaigns, which changes $\eta$ by less
than 0.2\% and the masses by at most $\pm$5\%.} We quote in
Table~\ref{tab:events} the energies for a minimum model value
$\eta=5.0\times10^{-4}$ as well. The impactor mass and diameter follow
from
\begin{equation}
m = \frac{2E_{k}}{v^{2}}, \qquad
D_{m} = 2\left(\frac{3m}{4\pi\rho}\right)^{1/3},
\label{eq:mass}
\end{equation}
with the Geminid bulk density $\rho = 3000$~kg\,m$^{-3}$\revR{, rounded from
$2900\pm600$~kg\,m$^{-3}$ ($2.9\pm0.6$~g\,cm$^{-3}$)} \citep{Babadzhanov2009}. Because $D_{m}\propto\rho^{-1/3}$, this choice,
at the upper end of the 1000--3000~kg\,m$^{-3}$ range measured for
Geminid meteoroids, makes the quoted diameters conservative: for
$\rho=2000$~kg\,m$^{-3}$ they would be 14\,\% larger and for
Phaethon's bulk density of $\sim$1700~kg\,m$^{-3}$ 21\,\% larger,
whereas the masses, fluxes, and all slope indices are independent of
$\rho$.

\subsection{Shower association}
\label{sec:assoc}

No other major shower was active around the observing epochs, so each
flash is either a Geminid or a sporadic impact. The Geminid probability
of each flash was quantified as
\begin{equation}
p = \frac{N^{\mathrm{ST}}}{N^{\mathrm{ST}} + N^{\mathrm{SPO}}},
\label{eq:pparam}
\end{equation}
where the shower and sporadic impact rates per unit lunar area scale as
$N^{\mathrm{ST}} \propto \revY{v^{\mathrm{ST}}} \gamma^{\mathrm{ST}}
\cos\varphi\, \sigma\, \mathrm{ZHR}^{\mathrm{ST}}_{\oplus}$ and
$N^{\mathrm{SPO}} \propto \revY{v^{\mathrm{SPO}}} \gamma^{\mathrm{SPO}}
\mathrm{HR}^{\mathrm{SPO}}_{\oplus}$, following the treatment of
\citet{Ortiz2015} and \citet{Madiedo2018}. Here \revY{$v^{\mathrm{ST}}$ and $v^{\mathrm{SPO}}$} are the impact
velocities ($35$ and $10$~km\,s$^{-1}$ for Geminid and sporadic
impactors), $\gamma$ the gravitational-focusing factors (0.915 and
0.773), $\varphi$ the \revY{incidence angle of the meteoroid,} measured from the local surface normal \revY{(the impact angle measured from the horizontal is $\theta=90\degr-\varphi$)}
(computed for each flash from its selenographic coordinates, the IAU
lunar rotation model including libration \citep{Archinal2018}, and the
Geminid radiant at $\alpha=112\degr$, $\delta=+32\degr$ corrected for the
orbital velocity of the Moon), $\sigma$ the
ratio of the Earth and Moon distances to the stream centre
($\sigma\simeq1$), and
$\mathrm{HR}^{\mathrm{SPO}}_{\oplus}=10$~h$^{-1}$ the mean sporadic
hourly rate. The Geminid rate
$\mathrm{ZHR}^{\mathrm{ST}}_{\oplus}$ at the epoch of each flash was
evaluated from the observed maximum rate of the corresponding return
and the post-maximum decline of the activity profile,
$\mathrm{ZHR}=\mathrm{ZHR}_{\max}10^{-B(\lambda_{\sun}-\lambda_{\max})}$
with the exponential slope $B=0.81$ of \citet{Jenniskens1994}: for 2018
from $\mathrm{ZHR}_{\max}=155.3$ at $\lambda_{\max}=262.358\degr$
\citep{Molau2020}, and for 2025 from $\mathrm{ZHR}_{\max}=142.8$ at the
nominal maximum $\lambda_{\max}=262.2\degr$. The solar longitude of each
flash was corrected for the displacement of the Moon along the orbital
motion of the Earth, which made the Moon cross a given stream
cross-section 3.7~h later than the Earth in 2018 and 3.1~h earlier in
2025; the resulting ``Moon-corrected'' solar longitudes are
$262.896\degr$--$263.028\degr$ (ZHR $=45$--57) in 2018 and
$262.825\degr$--$262.894\degr$ (ZHR $=39$--45) in 2025. The parameters \revY{$v$} and $\gamma$ of Eq.~(\ref{eq:pparam}) are those of
the earlier analyses and make $p$ essentially a measure of the ratio of
the visual shower and sporadic rates weighted by the impact geometry
\revV{(the velocities enter only as this conventional weighting and are not the
impact speed of 33.9~km\,s$^{-1}$ used for the energies in Sect.~\ref{sec:energy})}. \revV{An alternative, mass-based prescription gives lower values,
$p=0.34$--0.66 (Appendix~\ref{app:crater}).} The values of $p$ listed in Table~\ref{tab:events} are
those of Eq.~(\ref{eq:pparam}); the measured excess of the flux over the
sporadic background at the same masses (Sect.~\ref{sec:flux}) provides the
independent, population-level test of the shower association.

\subsection{Crater scaling}
\label{sec:crater}

\revV{The rim diameters of the craters expected in lunar regolith were
estimated for all events with the energy scaling relation} \revY{of \citet{Gault1974}, as compiled by} \citet{Melosh1989},
\begin{equation}
D_{c} = 0.25\,\rho_{p}^{1/6}\rho_{t}^{-1/2} E_{k}^{0.29}
(\sin\revY{\theta})^{1/3},
\label{eq:crater}
\end{equation}
with projectile density $\rho_{p}=3000$~kg\,m$^{-3}$, regolith density
$\rho_{t}=1500$~kg\,m$^{-3}$, and impact angle \revY{measured from the horizontal}
\revY{$\theta$}$=90\degr-\varphi$. \revV{These are the values listed in Table~\ref{tab:events}.} \revV{The $\pi$-group scaling of \citet{Holsapple1993} gives rim diameters
13--18\% smaller (Appendix~\ref{app:crater}).}

\section{Results}

\subsection{The 2018 events}
\label{sec:res2018}

The upper part of Table~\ref{tab:events} summarises the eleven 2018
flashes. The peak magnitudes span $R=5.88$--$9.52$ and the \revY{incidence} angles
$\varphi=\revP{11\degr}$--$\revP{70\degr}$. All events have Geminid association
probabilities $p=\revP{0.87}$--$0.96$; even halving the adopted ZHR leaves the
lowest value (flash H, $\varphi=\revP{70.1\degr}$) at $p=\revP{0.77}$, so the
association is robust against the uncertainty of the post-maximum
activity profile. It is also consistent with the simultaneous
multi-station confirmation of all eleven events. With
$\eta=1.4\times10^{-3}$--$5.0\times10^{-4}$ the impact energies span
$6.4\times10^{6}$--$5.1\times10^{8}$~J, corresponding to meteoroid masses
of \revN{11--890}~g, diameters of \revN{1.9--8.3}~cm, and predicted crater rim diameters
of 2.3--6.7~m. Blackbody temperatures of 2000--4000~K were derived for
these events from simultaneous low-dispersion spectroscopy
\citep{Yanagisawa2021}. \revV{The flashes occurred more than half a day after the visual maximum
(Sect.~\ref{sec:assoc}),}
a timing imposed by the lunar phase rather than by the shower; the
position of both campaigns in the activity profile is discussed in
Sect.~\ref{sec:stable}.

\begin{sidewaystable*}
\caption{Lunar impact flashes of the 2018 and 2025 Geminids observed at Nihon University.}
\label{tab:events}
\centering
\footnotesize
\setlength{\tabcolsep}{4pt}
\begin{tabular}{lcccccccccccc}
\toprule
Flash & Time & Lon. & Lat. & $R$ & $\varphi$ & $p$ &
\multicolumn{2}{c}{$E_{k}$ (J)} &
\multicolumn{2}{c}{Mass (g)} &
$D_{m}$ (cm) & $D_{c}$ (m) \\
\cmidrule(lr){8-9}\cmidrule(lr){10-11}
ID & (UT) & (deg) & (deg) & (mag) & (deg) & &
$\eta_{\mathrm{S11}}$ & $\eta_{\min}$ & $\eta_{\mathrm{S11}}$ & $\eta_{\min}$ & & \\
\midrule
\multicolumn{13}{c}{\itshape 2018 December 15 (UT)}\\
\midrule
B & 08:29:34 & \revP{$-56.8\pm0.2$} & \revP{$+31.2\pm0.2$} & $6.52\pm0.08$ & \revP{24.4} & 0.96 & $1.01\times10^{8}$ & $2.83\times10^{8}$ & \revN{175.8} & \revN{492.4} & \revN{4.82--6.79} & \revP{4.98--6.71} \\
C & 08:58:49 & \revP{$-51.1\pm0.2$} & \revP{$+38.0\pm0.2$} & $9.35\pm0.25$ & \revP{32.7} & 0.95 & $7.45\times10^{6}$ & $2.09\times10^{7}$ & \revN{13.0} & \revN{36.3} & \revN{2.02--2.85} & \revP{2.28--3.07} \\
D & 09:09:47 & $-57.9\pm0.2$ & \revP{$+0.4\pm0.2$} & $9.25\pm0.18$ & \revP{14.1} & 0.96 & $8.17\times10^{6}$ & $2.29\times10^{7}$ & \revN{14.2} & \revN{39.8} & \revN{2.08--2.94} & 2.45--3.31 \\
E & 09:44:04 & \revP{$-51.6\pm0.2$} & \revP{$-23.4\pm0.2$} & $8.30\pm0.13$ & \revP{37.0} & \revP{0.94} & $1.96\times10^{7}$ & $5.49\times10^{7}$ & \revN{34.1} & \revN{95.6} & \revN{2.79--3.93} & \revP{2.96--3.99} \\
G & 10:23:07 & \revP{$-44.1\pm0.2$} & \revP{$-9.7\pm0.2$} & $8.07\pm0.11$ & 31.6 & 0.95 & $2.43\times10^{7}$ & $6.79\times10^{7}$ & \revN{42.2} & \revN{118.2} & \revN{3.00--4.22} & 3.22--4.34 \\
H & 10:25:42 & \revP{$-84.7\pm1.9$} & \revP{$-59.7\pm0.3$} & $5.88\pm0.08$ & \revP{70.1} & \revP{0.87} & $1.82\times10^{8}$ & $5.11\times10^{8}$ & \revN{317.3} & \revN{888.6} & \revN{5.87--8.27} & \revP{4.25--5.73} \\
I & 10:28:46 & \revP{$-48.6\pm0.2$} & \revP{$-7.8\pm0.2$} & $6.96\pm0.08$ & \revP{27.0} & 0.95 & $6.74\times10^{7}$ & $1.89\times10^{8}$ & \revN{117.4} & \revN{328.7} & \revN{4.21--5.94} & 4.40--5.93 \\
J & 10:35:56 & \revP{$-65.0\pm0.2$} & \revP{$+28.3\pm0.2$} & $9.52\pm0.20$ & \revP{19.4} & 0.95 & $6.38\times10^{6}$ & $1.79\times10^{7}$ & \revN{11.1} & \revN{31.1} & \revN{1.92--2.71} & \revP{2.26--3.05} \\
K & 10:54:34 & \revP{$-64.3\pm0.2$} & \revP{$+24.4\pm0.2$} & $8.35\pm0.13$ & \revP{15.9} & 0.95 & $1.88\times10^{7}$ & $5.25\times10^{7}$ & \revN{32.7} & \revN{91.5} & \revN{2.75--3.88} & 3.11--4.19 \\
L & 11:22:12 & \revP{$-74.1\pm0.5$} & \revP{$-1.2\pm0.2$} & $7.42\pm0.10$ & \revP{11.3} & 0.95 & $4.43\times10^{7}$ & $1.24\times10^{8}$ & \revN{77.0} & \revN{215.7} & \revN{3.66--5.16} & \revP{4.02--5.41} \\
M & 11:35:52 & \revP{$-31.1\pm0.2$} & \revP{$+22.9\pm0.2$} & $7.04\pm0.10$ & \revP{39.7} & \revP{0.93} & $6.29\times10^{7}$ & $1.76\times10^{8}$ & \revN{109.4} & \revN{306.3} & \revN{4.11--5.80} & \revP{4.10--5.53} \\
\midrule
\multicolumn{13}{c}{\itshape 2025 December 14 (UT; 15 December JST)}\\
\midrule
A & 19:14:08 & \revP{$+79.3\pm1.0$} & \revP{$+7.1\pm0.3$} & $8.20\pm0.13$ & \revP{9.5} & 0.95 & $2.21\times10^{7}$ & $6.19\times10^{7}$ & \revN{38.5} & \revN{107.7} & \revN{2.90--4.09} & 3.29--4.44 \\
B & 19:31:27 & \revP{$+53.1\pm0.2$} & \revP{$-27.7\pm0.2$} & $8.64\pm0.10$ & \revP{51.5} & 0.92 & $1.47\times10^{7}$ & $4.12\times10^{7}$ & \revN{25.6} & \revN{71.8} & \revN{2.54--3.57} & \revP{2.51--3.38} \\
C & 19:35:55 & \revP{$+53.0\pm0.4$} & \revP{$+52.4\pm0.2$} & $9.61\pm0.20$ & \revP{49.5} & \revP{0.92} & $6.02\times10^{6}$ & $1.69\times10^{7}$ & \revN{10.5} & \revN{29.4} & \revN{1.88--2.65} & \revP{1.96--2.65} \\
D & 20:07:22 & \revP{$+29.2\pm0.2$} & \revP{$-23.6\pm0.2$} & $7.84\pm0.08$ & \revP{66.8} & \revP{0.87} & $3.07\times10^{7}$ & $8.59\times10^{7}$ & \revN{53.4} & \revN{149.6} & \revN{3.24--4.57} & \revP{2.67--3.60} \\
E & 20:09:51 & \revT{$+74.5\pm1.0$} & \revT{$+17.2\pm0.5$} & $9.69\pm0.19$ & \revT{13.7} & 0.94 & $5.58\times10^{6}$ & $1.56\times10^{7}$ & \revN{9.7} & \revN{27.2} & \revN{1.84--2.59} & \revT{2.20--2.96} \\
F & 20:18:46 & \revP{$+69.5\pm0.3$} & \revP{$+16.1\pm0.2$} & $9.12\pm0.12$ & \revP{17.9} & 0.94 & $9.44\times10^{6}$ & $2.64\times10^{7}$ & \revN{16.4} & \revN{46.0} & \revN{2.19--3.08} & \revP{2.54--3.42} \\
G & 20:31:40 & \revP{$+44.1\pm0.2$} & \revP{$+35.1\pm0.2$} & $7.63\pm0.08$ & \revP{45.6} & 0.92 & $3.72\times10^{7}$ & $1.04\times10^{8}$ & \revN{64.7} & \revN{181.3} & \revN{3.45--4.87} & \revP{3.41--4.60} \\
H & 20:33:08 & \revP{$+40.1\pm0.2$} & \revP{$-10.4\pm0.2$} & $8.89\pm0.12$ & \revP{51.6} & 0.91 & $1.17\times10^{7}$ & $3.26\times10^{7}$ & \revN{20.3} & \revN{56.8} & \revN{2.35--3.31} & \revP{2.34--3.16} \\
I & 20:52:55 & \revP{$+33.7\pm0.2$} & \revP{$+11.9\pm0.2$} & $8.69\pm0.11$ & \revP{52.0} & 0.91 & $1.40\times10^{7}$ & $3.92\times10^{7}$ & \revN{24.4} & \revN{68.2} & \revN{2.49--3.52} & \revP{2.46--3.32} \\
\bottomrule
\end{tabular}
\tablefoot{Flash IDs follow \citet{Yanagisawa2021} for 2018 and this
work for 2025 (the two ID sequences are independent). Times are the
start of the peak 16.67~ms frame. $R$ is the calibrated peak $R$-band
magnitude of that frame with its total uncertainty (Sect.~\ref{sec:phot});
$\varphi$ is the \revY{incidence} angle from the surface normal; $p$ is the Geminid
association probability of Eq.~(\ref{eq:pparam}) with the ZHR at the
Moon-corrected solar longitude of each flash (Sect.~\ref{sec:assoc}). All physical
quantities of both campaigns are computed with the same pipeline
(Sect.~\ref{sec:energy}) from the peak $R$ magnitudes, the topocentric distance at the
time of each flash, and \revN{$v=33.9$}~km\,s$^{-1}$: the two energy and mass
columns correspond to the velocity-dependent luminous efficiency of
\citet{Swift2011}, as adopted by \citet{Suggs2014},
$\eta_{\mathrm{S11}}=1.5\times10^{-3}\exp(-9.3^{2}/v^{2})=1.4\times10^{-3}$
at the Geminid lunar impact speed \revN{$v=33.9$}~km\,s$^{-1}$ (Eq.~\ref{eq:eta}),
and to the minimum model value $\eta_{\min}=5.0\times10^{-4}$, and the
ranges of
the meteoroid diameter $D_{m}$ (Eq.~\ref{eq:mass}) and crater rim
diameter $D_{c}$ (Eq.~\ref{eq:crater}) span the same interval. The
$\pm1\sigma$ magnitude uncertainty corresponds to a further
$\pm$8--25\% in mass. The 2018 peak magnitudes are measured on the
direct images at maximum light and therefore differ slightly from the
spectrally integrated magnitudes of \citet{Yanagisawa2021}\revY{ (Fig.~\ref{fig:uec})}.
\revP{Selenographic coordinates are measured in this work by registering the
earthshine-lit surface of each clip on the LROC WAC mosaic
(Sect.~\ref{sec:area}); for 2025~E we adopt the
position measured on the simultaneous Seimei/TriCCS image
(0.35\arcsec\ pixels; \citealt{Arima2026}), which our registration of the
Funabashi clip reproduces to within $0.3\degr$.} In our reduction of the simultaneous Hiratsuka 3~ms video,
flashes A and D are detected, B and H are marginal (S/N 8--9.5), C is
not detected (3$\sigma$ limit $R>9.1$ at 16.67~ms equivalent), and I is
detected with a second Hiratsuka camera at 100~fps; E, F, and G have no
simultaneous Hiratsuka clip, but E is confirmed by its simultaneous
detection with the 3.8~m Seimei telescope \citep{Arima2026}. Flashes C, E, and F are marginal detections
(S/N 8--12) in the Funabashi data.}
\end{sidewaystable*}

\subsection{The 2025 events}
\label{sec:res2025}

The lower part of Table~\ref{tab:events} lists the nine 2025 flashes.
All events have Geminid association probabilities $p=\revP{0.87}$--$0.95$
($p\ge\revP{0.77}$ with the ZHR halved), and every event was detected in at
least two consecutive video frames; the calibrated light curves of both
campaigns and the peak frame of every flash are shown in
Figs.~\ref{fig:app2018} and \ref{fig:app2025} (Appendix~\ref{app:flashes}). The peak magnitudes
span $R=7.63$--$9.69$ and the \revY{incidence} angles
$\varphi=\revP{9\degr}$--$\revP{67\degr}$. For the luminous-efficiency interval
$\eta=1.4\times10^{-3}$--$5.0\times10^{-4}$ the impact energies are
$5.6\times10^{6}$--$1.0\times10^{8}$~J, giving masses of \revN{10--181}~g,
diameters of 1.8--\revN{4.9}~cm, and predicted crater rim diameters of
\revP{2.0--4.6}~m, comfortably within the population observed in 2018. \revV{The
exponential decay times of the twelve flashes that lasted at least three
frames are $\tau=13$--54~ms and correlate with the peak magnitude rather
than with the mare or highland nature of the target
(Appendix~\ref{app:terrain}), and the simultaneous 3~ms photometry at
Hiratsuka shows that the flashes reach their maximum and fade by a factor
of several within the first few milliseconds (Appendix~\ref{app:phot}).}

\subsection{Magnitude distribution and population index}
\label{sec:magdist}

\begin{figure}
\centering
\includegraphics[width=\hsize]{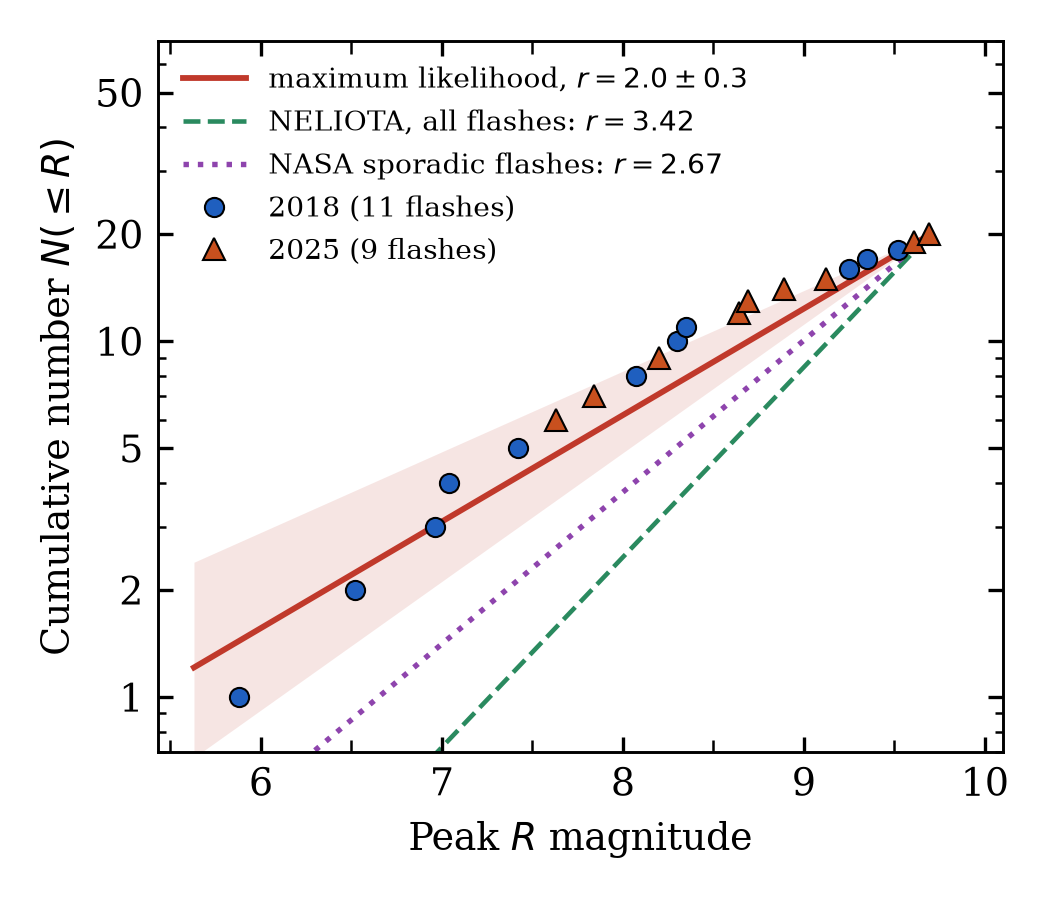}
\caption{Cumulative distribution of the peak $R$ magnitudes of the
combined 2018+2025 sample. Each symbol is one flash (circles: 2018;
triangles: 2025). Solid line and band: maximum-likelihood population
index $r=2.0\pm0.3$ for $R\le9.7$, drawn through the total number of
flashes at the limiting magnitude. Dashed and dotted lines: slopes of the
NELIOTA all-flash sample \citep[$r=3.42$;][]{Liakos2020} and of the NASA
sporadic sample \citep[$r=2.67$;][]{Suggs2014}, drawn from the same point.}
\label{fig:cummag}
\end{figure}

The peak-magnitude distributions of the two campaigns
(Fig.~\ref{fig:cummag}) are statistically consistent: the 2025 sample
(7.6--9.7~mag) overlaps the fainter part of the 2018 sample
(5.9--9.5~mag), the 2018 run having caught three flashes brighter than
$R=7$ in a 1.5 times longer exposure. Both samples terminate at $R\simeq9.5$--9.7, \revV{the faint-end cutoff of the
detected flashes (Sect.~\ref{sec:phot})}. Because the two campaigns
were obtained with the same telescope, comparable limiting magnitudes,
and similar Earth--Moon distances, the samples can be merged directly.
The cumulative $R$-magnitude distribution of the combined 20-flash
sample (Fig.~\ref{fig:cummag}) gives a population index
$r=2.0\pm0.3$ for $R\le9.7$ (maximum-likelihood estimate for an
exponential magnitude distribution truncated at the detection limit,
$\ln r=n/\sum_{i}(R_{\lim}-R_{i})$, with uncertainty $r\ln r/\sqrt{n}$;
a least-squares fit to the cumulative distribution gives $r=2.09$ with a
formal error of 0.07, a fit to 0.5-mag bins $2.3\pm0.2$, and
least-squares values of 1.8 and 2.6 for the 2018 and 2025 samples
alone). This is significantly ($>$4$\sigma$)
shallower than the NELIOTA all-flash value $r=3.42\pm0.22$
\citep{Liakos2020} and the NASA shower value $r=3.65\pm0.40$, and
shallower than, or at most comparable to, the NASA sporadic value
$r=2.67\pm0.24$ \citep{Suggs2014}, whose bright end is itself dominated
by large impactors. Bright flashes -- that is, large meteoroids -- are
therefore relatively overabundant in the Geminid stream, which we
interpret as a signature of a shower originating from an active
asteroid.

\subsection{Combined flux, mass, and size distributions}
\label{sec:flux}

With the monitored areas of Sect.~\ref{sec:area} and effective durations of
\revN{3.33} and 2.17~h, the 2018 campaign (eleven flashes over
$7.8\times10^{6}$~km$^{2}$) gives an impact flux of Geminid
meteoroids ($m\ga10$~g) of \revN{$4.2\times10^{-7}$}~km$^{-2}$\,h$^{-1}$ per unit
lunar surface and a mass-flux density of \revN{$3.6\times10^{-5}$}~g\,km$^{-2}$\,h$^{-1}$
($\eta=1.4\times10^{-3}$; $1.0\times10^{-4}$ for $\eta=5\times10^{-4}$).
For comparison, the eight 2018 flashes with $m\ge30$~g give
\revN{$3.1\times10^{-7}$}~km$^{-2}$\,h$^{-1}$, 3--5 times the sporadic
background flux of $\ge$30~g meteoroids,
$6.1\times10^{-10}$~m$^{-2}$\,yr$^{-1}$ $=7.0\times10^{-8}$~km$^{-2}$\,h$^{-1}$,
derived from the NASA lunar monitoring programme by \citet{Suggs2014}
(\revN{4.4} from the direct count, \revN{2.6} when the fitted distribution is scaled
from 10 to 30~g). The 2025 campaign
(nine flashes over $8.1\times10^{6}$~km$^{2}$) gives
\revP{$5.2\times10^{-7}$}~km$^{-2}$\,h$^{-1}$ and $1.5\times10^{-5}$~g\,km$^{-2}$\,h$^{-1}$ --
the same number flux within the Poisson uncertainties ($\pm30$--$35$\%),
the factor of \revN{2.4} lower mass flux reflecting the absence of flashes
brighter than $R=7.6$ in the shorter 2025 run. Referred to the collecting
area perpendicular to the Geminid radiant, which is the quantity that
describes the stream itself, the fluxes are \revN{$6.0\times10^{-7}$} and
$8.6\times10^{-7}$~km$^{-2}$\,h$^{-1}$ (combined \revP{$7.0\times10^{-7}$}); the larger
difference between the two years in this quantity reflects the more
oblique impact geometry of 2025 (mean $\varphi=\revP{40\degr}$ against $29\degr$ in 2018).

\begin{figure*}
\centering
\includegraphics[width=0.55\textwidth]{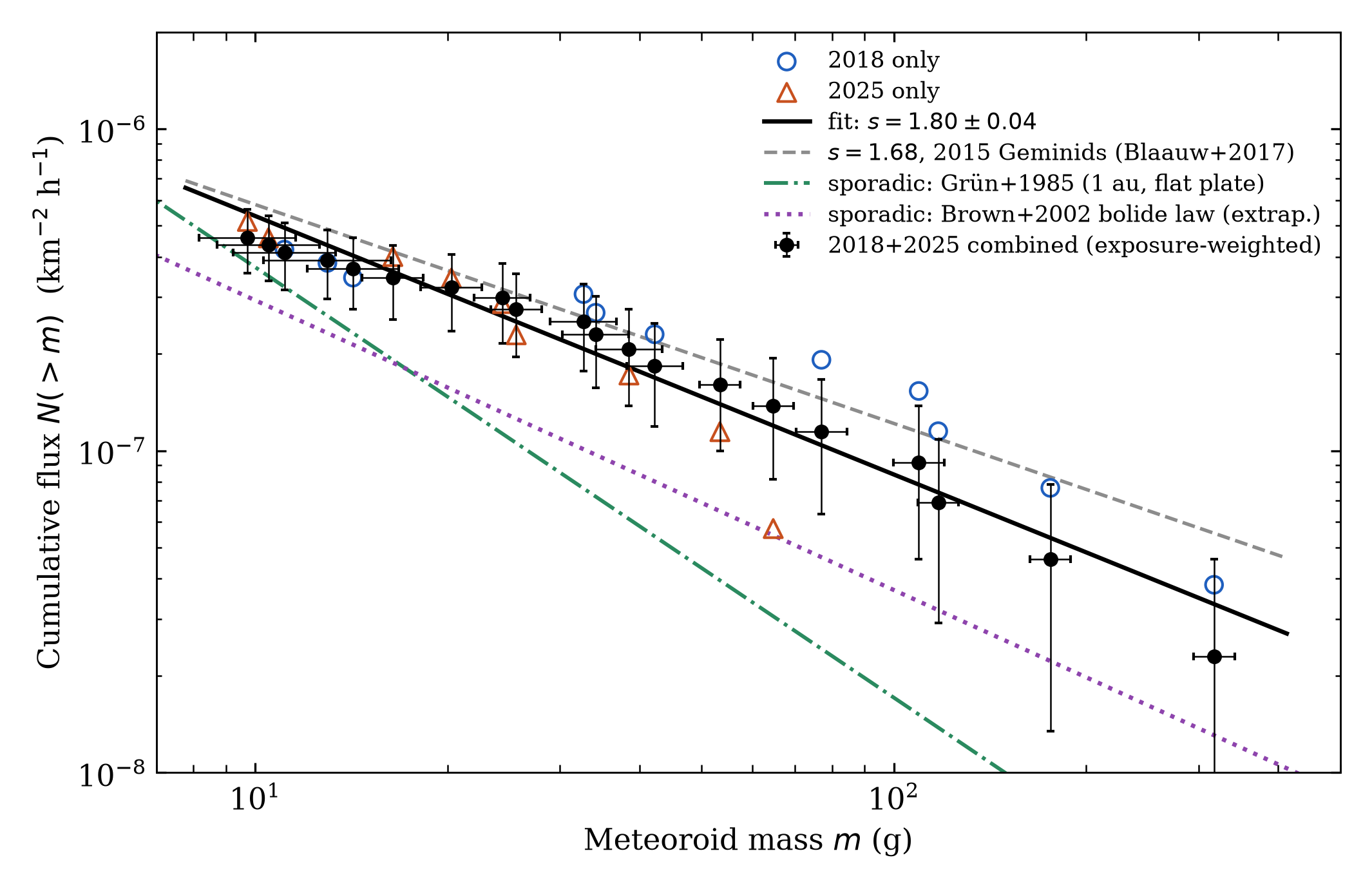}
\caption{Cumulative flux of Geminid meteoroids on the Moon as a function
of meteoroid mass ($\eta=1.4\times10^{-3}$, \revN{$v=33.9$}~km\,s$^{-1}$). Open
circles and triangles: 2018 and 2025 campaigns; filled circles: merged
sample, with its least-squares power-law fit (solid line, $s=1.80$).
Dashed line: slope $s=1.68$ of the 2015 Geminids \citep{Blaauw2017},
anchored at $m=30$~g. Dash-dotted and dotted lines: time-averaged
sporadic background per unit surface area from the interplanetary flux
model of \citet{Grun1985} at 1~au (extrapolated beyond its nominal 100~g
limit) and from the bolide power law of \citet{Brown2002}
(Sect.~\ref{sec:flux}).}
\label{fig:cummass}
\end{figure*}

\begin{figure*}
\centering
\includegraphics[width=0.82\textwidth]{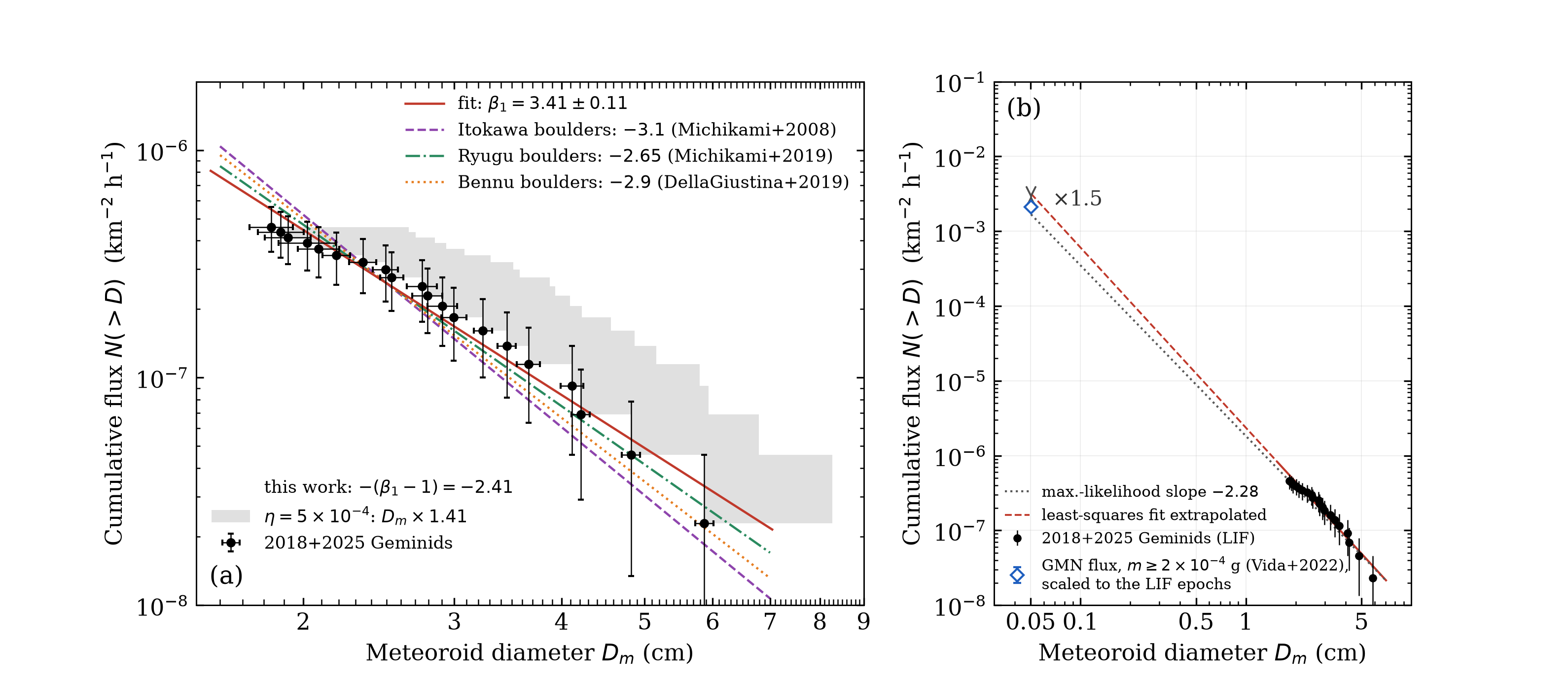}
\caption{Cumulative flux of the combined 2018+2025 sample as a function
of meteoroid diameter $D_{m}$. (a) Data (points) and least-squares
power-law fit of Eq.~(\ref{eq:sizeindex}) (line; cumulative index
$-2.41$). Horizontal bars: photometric uncertainty of the peak magnitude
propagated into $D_{m}$; vertical bars: Poisson uncertainty. Grey band:
shift of the distribution for $\eta=5\times10^{-4}$. Thin lines:
cumulative slopes of the boulder populations of Itokawa, Ryugu, and Bennu
\citep{Michikami2008,Michikami2019,DellaGiustina2019}, matched to the fit
at $D_{m}=2.5$~cm for a comparison of the slopes only. (b) The same data
and fit extrapolated (dashed) to the GMN limiting mass of
$2\times10^{-4}$~g ($D_{m}=0.5$~mm). Blue diamond: GMN optical flux of
the Geminids \citep{Vida2022} scaled to the LIF epochs and expressed per
unit lunar surface (bar: range between the two epochs). Dotted line:
maximum-likelihood slope $-2.3$ anchored at the smallest measured
diameter.}
\label{fig:cumsizeDm}
\end{figure*}

To exploit the full statistics we merged the two campaigns into a single
sample: the cumulative flux above a given mass is the number of flashes
above that mass in both campaigns divided by the summed exposure
$\sum_{y}A_{y}T_{y}=\revN{4.4}\times10^{7}$~km$^{2}$\,h, where $A_{y}$ and
$T_{y}$ are the monitored lunar area and effective duration of each
campaign (\revN{$2.6\times10^{7}$} and $1.75\times10^{7}$~km$^{2}$\,h for 2018
and 2025; the campaigns shown separately in the figures use their own
exposures). The
combined sample of 20 flashes yields a total flux of
\revN{$4.6\times10^{-7}$}~km$^{-2}$\,h$^{-1}$ and a mass-flux density of
$2.8\times10^{-5}$~g\,km$^{-2}$\,h$^{-1}$ ($\eta=1.4\times10^{-3}$).
Fitting cumulative power laws,
\begin{align}
\log_{10}N(>m) &= \log_{10}N_{1}-(s-1)\log_{10}m,\label{eq:massindex}\\
\log_{10}N(>D_{m}) &= \log_{10}N_{2}-(\beta_{1}-1)\log_{10}D_{m},\label{eq:sizeindex}\\
\log_{10}N(>D_{c}) &= \log_{10}N_{3}-(\beta_{2}-1)\log_{10}D_{c},\label{eq:craterindex}
\end{align}
to the combined sample by least squares yields $s=1.80$ over the full
mass range \revN{10--320}~g (Fig.~\ref{fig:cummass}), $\beta_{1}=3.41$, and
$\beta_{2}=\revP{3.90}$ (lines in Figs.~\ref{fig:cumsizeDm} and
\ref{fig:cumsizeDc}; formal errors 0.04, 0.11, and \revP{0.19}).

Because the
points of a cumulative distribution are strongly correlated, these
formal errors are not confidence intervals. We therefore adopt the
maximum-likelihood estimator for a power law truncated at the
completeness limit, $s-1=n/\sum_{i}\ln(m_{i}/m_{\min})$ with
uncertainty $(s-1)/\sqrt{n}$ \citep{Clauset2009}, which gives
$s=1.76\pm0.17$ for the 20 flashes and $1.75\pm0.17$ for the 19
flashes with $m\ge10$~g, the completeness limit implied by the
$R\simeq9.7$ limiting magnitude at the 2025 Earth--Moon distance.
Fitted separately, the campaigns give $s_{2018}=1.66\pm0.20$ ($n=11$)
and $s_{2025}=2.10\pm0.37$ ($n=9$) by maximum likelihood (1.63 and
2.04 by least squares), consistent with each other and with the
combined value within their uncertainties. The
size and crater-diameter indices follow from $s$ through the scaling
exponents of Eqs.~(\ref{eq:mass}) and (\ref{eq:crater}),
$\beta_{1}=1+3(s-1)=3.3\pm0.5$ and $\beta_{2}=1+(s-1)/0.29=3.6\pm0.6$
(direct maximum-likelihood fits give $3.3\pm0.5$ for both). In the
following we quote the maximum-likelihood values with their
uncertainties and keep the least-squares lines in the figures, which
lie within one standard deviation of them. The two
faintest flashes (2025~C and E, \revN{10}~g) lie at the limit and the slight
flattening of the cumulative curve there (Fig.~\ref{fig:cummass}) is an
incompleteness effect.

These values are consistent with the preliminary
2018-only indices ($s=1.78\pm0.16$, $\beta_{1}=3.23\pm0.28$,
$\beta_{2}=3.43\pm0.51$; \citealt{Abe2024}) and with $s=1.68\pm0.04$
derived for the 2015
Geminids over the mass range $1.8\times10^{-4}$--30~g from combined CMOR
radar, wide-field camera, and NASA lunar-flash observations
\citep{Blaauw2017}, extending the validity of a shallow Geminid mass
distribution up to \revV{$\sim$300~g}. The crater-diameter index connects,
within errors, to the production size-frequency distribution slope
$\beta\simeq3.6$ measured for lunar craters of 10~m--1~km
\citep{Morota2015}, bridging the gap below 10~m.

Figure~\ref{fig:cummass}
also compares the Geminid flux with the time-averaged sporadic background:
the interplanetary flux model of \citet{Grun1985} at 1~au, which is
\revN{1.4}--5 times lower than the Geminid peak
flux at 10--100~g and falls more steeply, and the bolide power law of
\citet{Brown2002} extrapolated downward by eight orders of magnitude in
energy, which lies a factor of $\sim$2 below the Geminid peak flux
(ratio 1.8--2.3) when expressed, like the LIF flux, per unit surface
area. \revV{(the conversion is given in Appendix~\ref{app:crater}).} The agreement with the \citet{Brown2002} law
extends to the slope: its cumulative energy index of 0.90 corresponds to
$s=1.90$, within one standard deviation of the Geminid value
$s=1.76\pm0.17$, so that a common power law can describe decimetre-to-metre bolides and
centimetre Geminid pebbles over eight orders of magnitude in mass. Such
a slope is characteristic of collisionally or thermally fragmented
material and matches the boulder populations of rubble-pile asteroids
(Sect.~\ref{sec:large}). \revV{The shower peak thus exceeds the time-averaged sporadic
background by a factor of only 1.4--5 at these masses, in line with the
sub-per-cent annual Geminid contribution derived in crater space below.}

\revV{Expressed as mass indices through $s=1+2.5\log_{10}r$, the reference
population indices of Sect.~\ref{sec:magdist} correspond to $s=2.34$
(NELIOTA all flashes), 2.41 and 2.07 (NASA shower and sporadic flashes),
and 2.04 (visual Geminids, $r=2.6$), whereas our $r=2.0\pm0.3$ gives
$s=1.75$, consistent with the direct estimate. The Geminid centimetre
population is thus flatter than every reference distribution except the
Geminids themselves at smaller masses \citep{Blaauw2017}.}

\revV{The difference between the two campaigns in Fig.~\ref{fig:cummass} is
largely a bright-end shortfall in the shorter 2025 run. It is not
statistically significant and accounts for the factor of 2.4 between the
two mass fluxes. The luminous efficiency rescales the whole mass axis as
$\eta^{-1}$ without changing the slopes (Appendix~\ref{app:crater}).}

\revV{In terms of craters (Fig.~\ref{fig:cumsizeDc}), the production rate of
2--5~m craters during the few hours of the Geminid maximum exceeds the
long-term mean production function by a factor of $\sim$20 ($\sim$10--50
when the uncertainties of the luminous efficiency and of the scaling law
are included), whereas the annual Geminid contribution to the metre-scale
cratering rate is of order half a per cent (Appendix~\ref{app:crater}).}

\section{Discussion}

\subsection{A stable centimetre-sized Geminid population}
\label{sec:stable}

The impact fluxes measured seven years apart --
\revN{$4.2\times10^{-7}$}~km$^{-2}$\,h$^{-1}$ in 2018 and
\revP{$5.2\times10^{-7}$}~km$^{-2}$\,h$^{-1}$ in 2025 -- agree within the
$\sim$30\% Poisson uncertainties, and the corresponding maximum visual
ZHRs (155 and 143) differ by less than 10\%. The cross-section of the
Geminid stream swept by the Earth--Moon system therefore contained an
essentially unchanged spatial density of centimetre-sized particles at
the two epochs. Dividing the combined flux by the lunar impact velocity,
and neglecting the modest ($\gamma\simeq0.9$) gravitational focusing of
the Earth--Moon system, the corresponding spatial number density of
$m\ga10$~g ($D_{m}\ga1.9$~cm) Geminids near the stream core at 1~au is
$n \simeq 4\times10^{-12}$~km$^{-3}$, i.e. of order one such particle
per cube of $\sim$6000~km on a side during the $\sim$3~h around the
peak. Stream models predict that the centimetre-sized
population, being least affected by radiation pressure and
Poynting--Robertson drag, traces the original ejection epoch most
faithfully \citep{Ryabova2012,Cukier2023}.

\begin{figure*}
\centering
\includegraphics[width=0.55\textwidth]{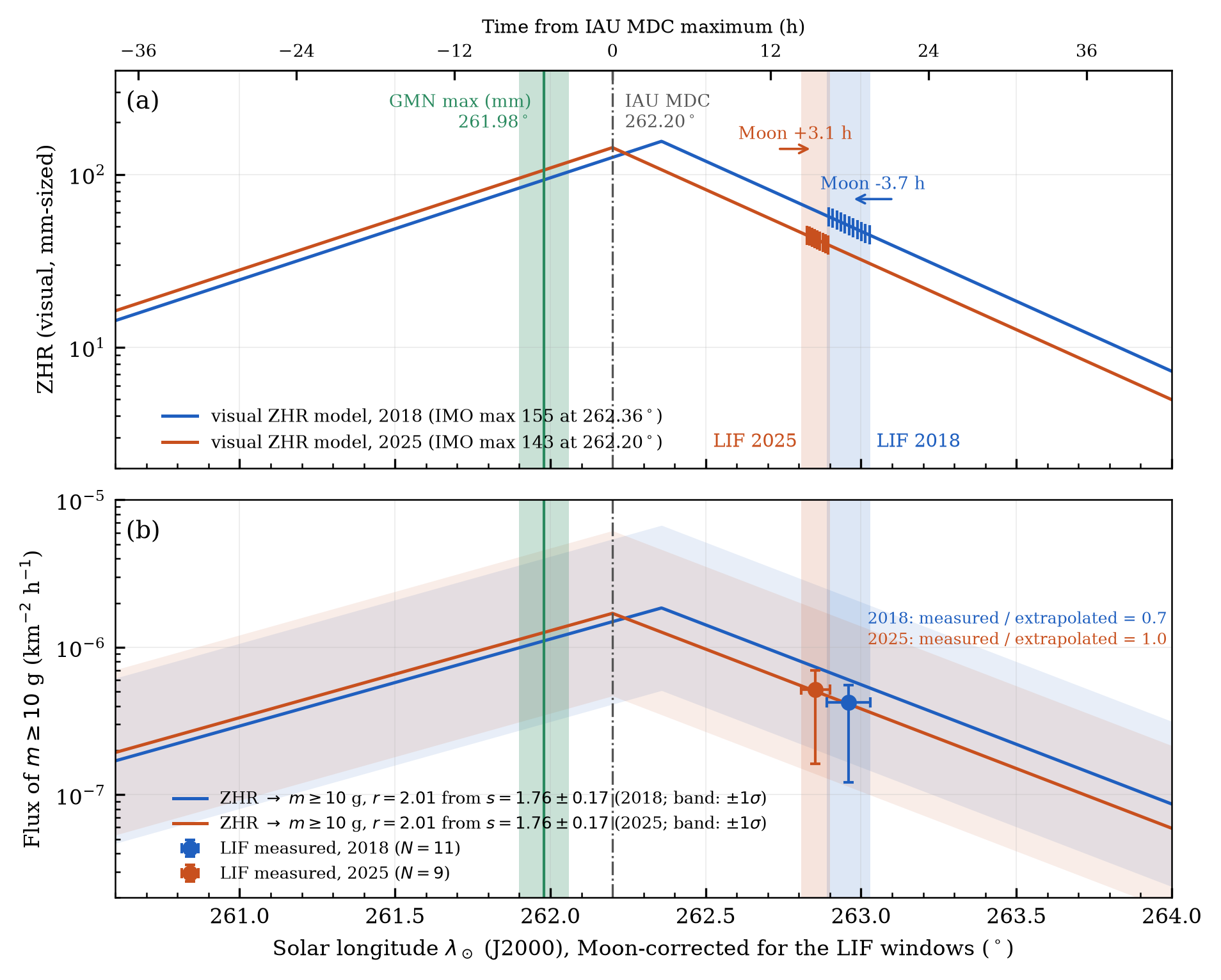}
\caption{LIF observing windows in the Geminid activity profile.
(a) Visual ZHR of the millimetre-sized population as a function of solar
longitude for 2018 (blue) and 2025 (red); the upper axis gives the time
from the IAU MDC maximum. Green: GMN optical-flux maximum with its
uncertainty \citep{Vida2022}; dash-dotted: IAU MDC maximum. Shaded bands:
LIF windows in Moon-corrected solar longitude; arrows: shift from the
Earth's to the Moon's position along the orbit; ticks: individual
flashes. (b) The same profiles converted to the flux of $m\ge10$~g
meteoroids per unit lunar surface with the mass index $s=1.76\pm0.17$
measured here (solid lines; bands: $\pm1\sigma$ range of $s$). Points:
fluxes measured by LIF (vertical bars: Poisson uncertainty; horizontal
bars: observing windows). The labels give the ratio of measured to
extrapolated flux.}
\label{fig:profile}
\end{figure*}

Table~\ref{tab:offset} and Fig.~\ref{fig:profile} place the two LIF
epochs in the stream. \revV{Panel (a) of Fig.~\ref{fig:profile} shows the
visual activity profile of the millimetre-sized population, modelled
with the two-component Geminid slopes of \citet{Jenniskens1994}
($B=0.59$ before and 0.81 after maximum) anchored to the IMO maxima of
the two returns (Sect.~\ref{sec:assoc}), together with the GMN and IAU MDC
maxima and the two LIF windows in Moon-corrected solar longitude.
Panel (b) converts the same profiles into
the flux of $m\ge10$~g meteoroids per unit lunar surface with the
extrapolation described in Table~\ref{tab:offset} and compares it with
the fluxes measured by the two campaigns.}

Referred to the maximum of the millimetre-sized population -- the IAU
Meteor Data Center value $\lambda_{\max}=262.2\degr$ \citep{Hajdukova2023},
the Global Meteor Network optical flux peak at $261.98\pm0.08\degr$
\citep{Vida2022}, or the IMO 2018 maximum at $262.358\degr$
\citep{Molau2020} -- the Moon sampled the stream
14--23~h (\revN{1.5}--2.5$\times10^{6}$~km along the
orbit) after the peak in 2018 and 12--21~h
(1.3--2.2$\times10^{6}$~km) after it in 2025, where the visual rate had
fallen to 16--40\% of its maximum. Both LIF windows were dictated by the
visibility of the lunar night side and do not bracket the maximum, so the
time of the centimetre-particle peak itself cannot be measured from these
data. 

\revV{What can be measured is the centimetre-sized flux at these two
points of the stream. It exceeds the extrapolation of the simultaneous
visual ZHR to 10~g by a factor of 6--8 if the visual population index
$r=2.6$ ($s=2.04$) is used, but agrees with it for the slopes measured
here: the ratio of measured to extrapolated flux is 1.0--1.4 for the
least-squares slope ($s=1.80$) and 0.7--1.0 for the maximum-likelihood
slope ($s=1.76$) (Fig.~\ref{fig:profile}b, Table~\ref{tab:offset}).}
\revV{The extrapolation spans eight magnitudes and is therefore sensitive to
the adopted index (band in Fig.~\ref{fig:profile}b;
Appendix~\ref{app:stream}).} \revV{The discrepancy with $r=2.6$ is therefore a statement about the population
index:} $r=2.6$, which describes the millimetre-sized meteors, does not
hold above $\sim$10~g, where the distribution is as shallow as
$s\simeq1.75$--1.8 (Sect.~\ref{sec:large}).

Whether, in addition, the
centimetre-sized activity profile is broader and peaks later than the
millimetre one -- as expected from radiation-pressure and
Poynting--Robertson sorting of the stream \citep{Liou1995,Ryabova2012},
the larger particles being the least displaced from the parent orbit
and populating the core that the Earth crosses after the dust-dominated
outer part -- cannot be decided from two post-maximum windows. A
decisive test requires LIF monitoring on both sides of the visual
maximum in the same return, which the lunar phase allows every few
years.

The measured stability and the shallow mass index together provide quantitative constraints for models of the
Phaethon--Geminid stream complex, whose proposed formation ages range
from $\sim$2~kyr to $\sim$18~kyr
\citep{Ryabova2012,Jo2024,MacLennan2024}, and complement the direct
imaging of the stream density core by Parker Solar Probe/WISPR
\citep{Battams2025}.

\subsection{Large particles from an active asteroid}
\label{sec:large}

The combined population index $r=2.0\pm0.3$ and mass index
$s=1.76\pm0.17$ of the Geminid flashes are markedly shallower than the
NELIOTA all-flash values \citep{Liakos2020,Liakos2024}, implying a
relative excess of particles of tens of grams -- several centimetres --
in the stream. The cumulative size index of the Geminid pebbles,
$-(\beta_{1}-1)=-2.3\pm0.5$ (maximum likelihood; $-2.4$ for the
least-squares fit), is close to the power indices of the metre-scale
boulder populations of the rubble-pile asteroids Ryugu ($-2.65$;
\citealt{Michikami2019}), Bennu ($-2.9$; \citealt{DellaGiustina2019}),
and Itokawa ($-3.1$; \citealt{Michikami2008}) (Fig.~\ref{fig:cumsizeDm}),
as expected if the pebbles are fragments of the same kind of
collisionally and thermally fractured material.

\revV{Three points qualify this comparison (Appendix~\ref{app:frag}). The
size index follows exactly from the mass index, $\beta_{1}-1=3(s-1)$, so
that Fig.~\ref{fig:cumsizeDm} is the size-space representation of $s$. The
comparison with the boulder populations spans more than two orders of
magnitude in size and tests the slope, not the level. Finally, the diurnal
thermal skin depth of Phaethon is 2--5~cm, which makes thermal-fatigue
fragmentation \citep{Delbo2014,Molaro2020} a natural source of fragments
of the measured size.}

\revV{The LIF power law can be tied directly to the millimetre-sized
population without passing through the visual ZHR
(Fig.~\ref{fig:cumsizeDm}b). Extrapolated from 1.9~cm to the GMN limiting
mass of $2\times10^{-4}$~g ($D_{m}=0.5$~mm), the maximum-likelihood slope
reproduces the GMN optical flux of the Geminids \citep{Vida2022}, scaled
to the LIF epochs, to within 30\,\%, and the least-squares slope to
within a factor of 1.4--1.7 (Appendix~\ref{app:frag}).} Equivalently, the
effective mass index between $2\times10^{-4}$~g and 10~g, $s=1.77$,
coincides with the LIF value $s=1.76\pm0.17$ and with the $1.68\pm0.04$
of \citet{Blaauw2017}: within the uncertainties a single power law with
$s\simeq1.75$--1.8 describes the Geminid mass distribution from
0.2~mg to 300~g. The apparent excess of centimetre-sized particles in
Sect.~\ref{sec:stable} is thus an excess relative to the visual
population index $r=2.6$ ($s=2.04$), not relative to the stream's own
shallow distribution, which the GMN and LIF data trace consistently.

\revV{The closest direct analogue of such ejecta is the particle-ejection
activity of the B-type near-Earth asteroid (101955) Bennu
\citep{Lauretta2019,Hergenrother2020}, whose ejected particles ($<$1 to
$\sim$10~cm) cover the size range of the Geminid impactors measured here.
Bennu-like activity, however, cannot have built the stream: its launch
speeds are an order of magnitude below the escape speed of Phaethon, and
its mass-loss rate of $\sim$10~kg\,yr$^{-1}$ falls short of the stream mass
\citep[at least $1.6\times10^{13}$~kg;][]{Blaauw2017} by many orders of
magnitude (Appendix~\ref{app:frag}). The centimetre-sized population
therefore records one or a few episodes of intense mass loss, consistent
with the short ejection intervals favoured by stream models
\citep{Ryabova2012,Jo2024}.} The
sodium-depletion measured in Geminid meteors \citep{Abe2020} and the
survival of large compact grains are both consistent with strong thermal
processing of the ejected material. LIF monitoring thus provides the
size-distribution anchor at centimetres that neither in-situ dust
detectors (\revS{sub-micrometre to} micrometre\revS{ sizes}, as planned for \revS{the Dust Analyser of} DESTINY$^{+}$;
\citealt{Ozaki2022,Arai2024}\revS{; \citealt{Simolka2024}}) nor visual/radar meteor networks
(sub-millimetre to millimetre) can supply.

\subsection{Towards a direct calibration of the luminous efficiency}

All mass estimates scale inversely with the luminous efficiency, which
is uncertain by an order of magnitude
($\eta\sim5\times10^{-4}$--$6\times10^{-3}$;
\citealt{BellotRubio2000,Ortiz2015,Fuse2020}, \revR{\citealt{Sheward2025}}). The predicted
crater rim diameters of the 2018 and 2025 events (2--7~m for the \revY{Gault}
scaling, 1.5--4~m for the $\pi$-group scaling) and the
\revP{impact coordinates measured here (uncertainties of typically $0.2\degr$,
i.e. 6~km, and up to $1\degr$--$2\degr$ in longitude near the limb)} make a search in Lunar
Reconnaissance Orbiter Narrow Angle Camera (NAC) temporal image pairs
feasible, following the approach that identified 222 new impact craters
\citep{Speyerer2016}. Each NAC pair images a $5\times25$~km strip
($125$~km$^{2}$) at 0.5~m~pixel$^{-1}$ from the 50~km mapping orbit
\citep{Robinson2010}, so a rim-to-rim diameter of 2--5~m spans only
4--10 pixels; a fresh crater is, however, surrounded by an ejecta and
reflectance disturbance several crater diameters across, which is what
allowed \citet{Speyerer2016} to identify new craters from several metres
up to 43~m in 14\,092 temporal pairs spanning seven years. Our predicted
sizes are thus at, not below, the demonstrated detection limit. The
search area per flash, \revP{$\pi(6$--$30~{\rm km})^{2}\simeq10^{2}$--$3\times10^{3}$~km$^{2}$,
corresponds to between one and about twenty NAC frame
pairs, so the flashes away from the limb, with
$0.2\degr$ positions, and the brightest events}
($D_{\rm c}\gtrsim4$~m) are the natural targets. Because temporal
pairs exist only where repeat NAC coverage under similar illumination
happens to be available, dedicated post-impact NAC imaging of the twenty
sites matched to the archival coverage is required; the 2025 sites are
particularly favourable, because the archive now provides more than
fifteen years of pre-impact images. Because the Geminid impact
velocity is known \revY{to within $\sim$1~km\,s$^{-1}$ (Sect.~\ref{sec:energy})}, the identification of even a single flash
crater would eliminate the velocity--mass degeneracy and calibrate $\eta$
directly at \revY{33--35}~km\,s$^{-1}$, anchoring laboratory results obtained at
lower velocities \citep{Fuse2020}. The located impacts also constitute
time-stamped seismic sources for future lunar seismometer networks
\citep{Yamada2011}.

\subsection{Future monitoring}

The consistency of two short campaigns demonstrates that even
moderate-aperture suburban systems contribute meaningful statistics
when flashes are validated by multi-station coincidence. Extending the
temporal coverage -- through networked stations at different longitudes,
daytime short-wave-infrared observations \citep{Sheward2024}, high-speed
photometry resolving the flash rise \citep{Giancono2026}, and ultimately
continuous monitoring from cislunar platforms \citep{Funase2020} -- will
refine the flux and its year-to-year variation, of direct relevance both
to Phaethon science ahead of the DESTINY$^{+}$ flyby and to impact-risk
assessment for lunar surface assets.

\section{Conclusions}

We have presented a homogeneous analysis of lunar impact flash
observations obtained with the 0.40~m GUNDAM telescope at Nihon
University
at the maxima of the 2018 and 2025 Geminid meteor showers. Eleven flashes
were confirmed in \revN{3.3}~h in 2018 and nine flashes in $\sim$2.2~h in
2025, each validated by simultaneous detections at independent stations,
and every event has a Geminid association probability above 0.8
(\revP{0.87}--0.96 in 2018, \revP{0.87}--0.95 in 2025; \revP{0.3}--0.7 with a mass-based
prescription). The
two campaigns, with monitored night-side areas measured on the video
frames, yield mutually consistent impact fluxes of \revN{$4.2\times10^{-7}$} and
\revP{$5.2\times10^{-7}$}~km$^{-2}$\,h$^{-1}$, indicating a stable spatial density of
centimetre-sized Geminids in the stream over seven years; the merged
exposure-weighted sample gives a combined flux of
\revN{$4.6\times10^{-7}$}~km$^{-2}$\,h$^{-1}$, a mass-flux density of
$2.8\times10^{-5}$~g\,km$^{-2}$\,h$^{-1}$, and a spatial number density of
$\sim$$4\times10^{-12}$~km$^{-3}$ for $m\ga10$~g particles near the
stream core. For the nominal luminous efficiency $\eta=1.4\times10^{-3}$
the impactors have masses of $\sim$\revN{10--320}~g and diameters of
1.8--\revN{5.9}~cm, and should have produced regolith craters of 2--5~m rim
diameter (up to 7~m for the minimum luminous efficiency). The combined
sample of 20 flashes gives a maximum-likelihood mass index
$s=1.76\pm0.17$ (unchanged above the 10~g completeness limit), size
indices $\beta_{1}=3.3\pm0.5$ and $\beta_{2}=3.6\pm0.6$, and a
population index $r=2.0\pm0.3$, all shallower than the NELIOTA
all-flash background, consistent with the radar-to-lunar Geminid value
of \citet{Blaauw2017}, and indicative of an excess of large particles
ejected from the active asteroid (3200) Phaethon. Both campaigns
sampled the stream 0.5--1~d after the visual maximum, where the
centimetre-sized flux exceeds the extrapolation of the visual rate with
its population index $r=2.6$ by a factor of 6--8, whereas a single
shallow power law ($s\simeq1.75$--1.8) connects the Global Meteor
Network millimetre flux and the LIF centimetre flux at the same epoch;
measuring the centimetre-particle peak itself requires LIF coverage on
both sides of the maximum. Identification of the
predicted metre-class craters in LRO temporal imaging would provide the
first direct calibration of the luminous efficiency at a known impact
velocity, and the located impacts will serve as calibration sources for
future lunar seismic networks.

\section*{Data availability}

The reduced video clips of the twenty flashes, the calibrated light
curves, the event table (Table~\ref{tab:events}), and the analysis
scripts used in this paper \revF{are publicly available at
\mbox{\url{https://doi.org/10.5281/zenodo.24391537}} (Zenodo repository).}

\begin{acknowledgements}
The GUNDAM lunar-impact-flash telescope system was established with
the Leading Research Promotion Grant of the Research Institute of
Science and Technology, Nihon University, in cooperation with Kazuaki
Watanabe and Hideyuki Fujimura (Showa Kikai Seisakusho Co., Ltd.) and
Officina Stellare S.r.l.\revV{; S.S. led the consultations with
the manufacturers, the design of the dome, and the structural
calculations of the movable outrigger, and T.T. supported the
construction of the telescope. We thank Hitoshi Negoro and Tsukie
Izuhara (Nihon University) for their support of the GUNDAM project.}
We thank the past members of Abe's
laboratory who took part in the LIF observations with the GUNDAM
telescope: Yu Fukutomi, Takumi Sato, Shogo Nakamura, Hikaru Endo,
Ryota Fukazawa, Keisuke Udaka, and Sayaka Karube. The 2018 campaign was supported by the JSPS Bilateral Joint
Research Project with the participation of P.~Lognonn\'e (IPGP),
J.~Vaubaillon (Observatoire de Paris), C.~Avdellidou (Observatoire
de la C\^ote d'Azur)\revV{, and T.~Fukuhara (Rikkyo University)}. \revF{This work was supported by the 2026 Joint Research Program of the Aizu Research Center for Space Informatics (ARC-Space), University of Aizu.} We thank
the International Meteor Organization and its visual observers for the
Geminid activity profiles. \revV{An AI assistant (Claude, Anthropic) was used to check
the manuscript; the authors reviewed all content and take full
responsibility for it.}
\end{acknowledgements}


\begin{appendix}

\section{Independent photometric checks}
\label{app:phot}
\label{sec:hira}

\begin{figure}
\centering
\includegraphics[width=\hsize]{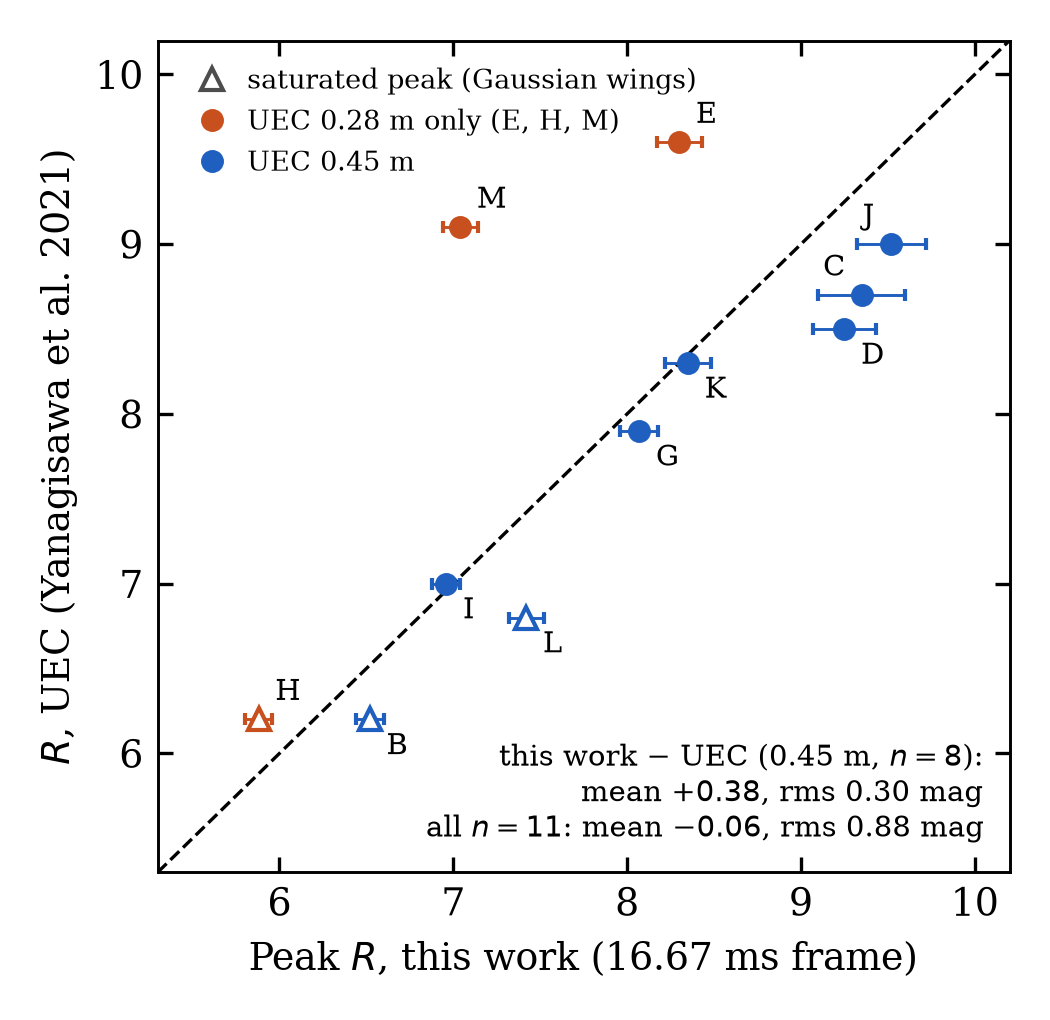}
\caption{Peak $R$ magnitudes of the eleven 2018 flashes measured in this
work (Funabashi) against those measured simultaneously at UEC
\citep[Table~2]{Yanagisawa2021}. \revY{Blue symbols: flashes recorded with
the UEC 0.45~m telescope; orange symbols: flashes E, H, and M, recorded
at UEC only with the 0.28~m telescope.} Triangles: flashes whose peak
frame saturated the 8-bit camera at Funabashi. The dashed line is the
identity.}
\label{fig:uec}
\end{figure}

\revV{The calibration of Sect.~\ref{sec:phot} is supported by two independent data
sets: the $R$ magnitudes measured simultaneously at UEC in 2018
(Fig.~\ref{fig:uec}; the mean difference, this work minus UEC, is
$+0.38$~mag for the eight flashes recorded with the 0.45~m telescope and
$-0.06$~mag with an rms scatter of 0.88~mag for all eleven) and the high-speed video obtained at Hiratsuka in
2025, described here.}

\begin{figure}
\centering
\includegraphics[width=\hsize]{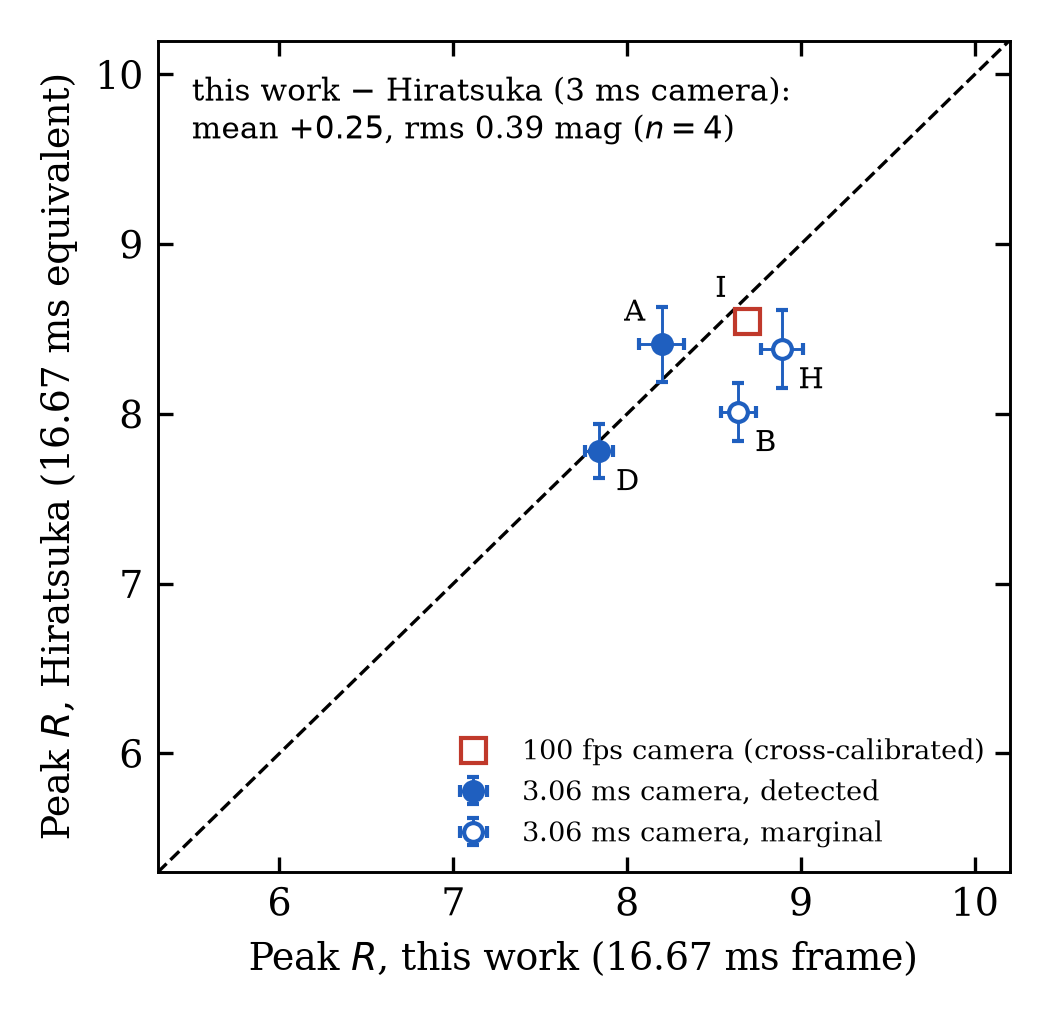}
\caption{Peak $R$ magnitudes of the 2025 flashes at Funabashi against
those at Hiratsuka (3.06~ms frames \revY{summed over the brightest
16.67~ms window}; open circles: marginal Hiratsuka detections). The open
square is flash I in the Hiratsuka 100~fps camera, whose zero point was
fixed on the time-integrated light of this flash. The dashed line is the
identity.}
\label{fig:fujii}
\end{figure}

\begin{figure*}
\centering
\includegraphics[width=0.9\textwidth]{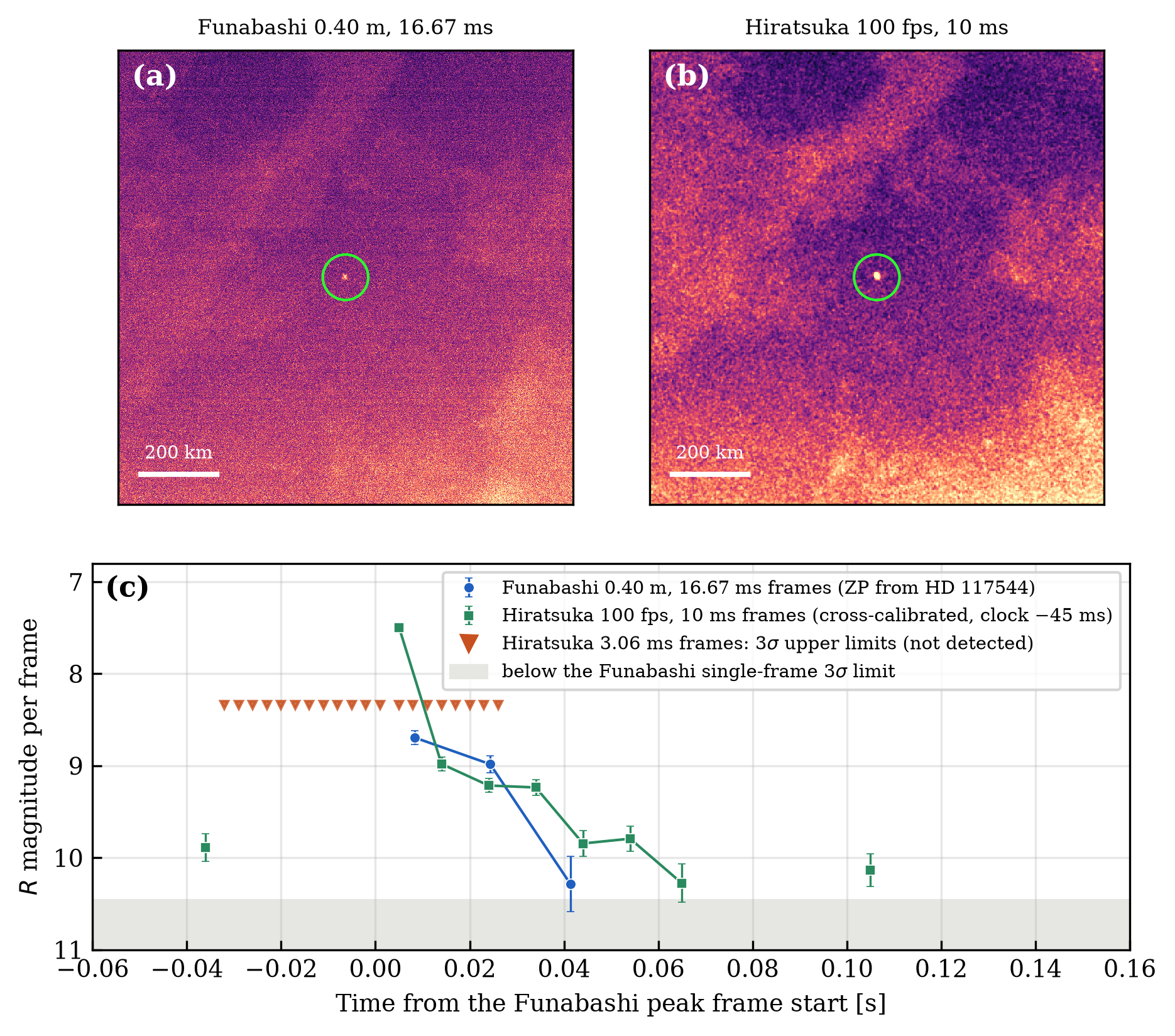}
\caption{Lunar impact flash I of the 2025 campaign (14 December 2025,
20:52:55.86~UT). (a) Peak 16.67~ms frame of the Funabashi 0.40~m
telescope. (b) Peak 10~ms frame of the Hiratsuka 100~fps camera. Each
cut-out covers $1200\times1200$~km on the Moon in the native orientation
of the camera; the circle marks the photometric aperture. (c) Calibrated
$R$-band light curves on a common time axis: Funabashi (blue; one bar per
16.67~ms frame, $1\sigma$ errors), Hiratsuka 100~fps (green; 10~ms
frames, clock offset of $-45$~ms applied), and $3\sigma$ upper limits of
the Hiratsuka 3.06~ms frames (triangles). The grey band lies below the
Funabashi single-frame $3\sigma$ limit.}
\label{fig:multicam}
\end{figure*}

The 2025 flashes were also recorded at Hiratsuka (Kanagawa), 70~km from
Funabashi, \revY{with three cameras}\revF{, whose recording computers were synchronised to an NTP server every 5~min}\revY{. (i) A monochrome CMOS camera}
\revF{(Sony IMX287 sensor, $720\times540$ pixels of $6.9\,\mu$m) on a 0.20~m Celestron C8 telescope with a HyperStar corrector (nominal focal length 390~mm, f/1.9), without a filter,} \revY{recorded
the whole lunar disc continuously at 326~frames~s$^{-1}$ with an exposure
of 3.06~ms per frame ($3.9\arcsec$~pixel$^{-1}$
from the radius of the lunar disc}\revF{; $3.65\arcsec$ for the nominal focal length}\revY{). (ii) A second} \revF{IMX287 camera on a 0.15~m Celestron C6 with a HyperStar corrector (focal length 300~mm, f/2.0) was}
\revY{triggered by motion detection at
270~frames~s$^{-1}$; its data were not used here. (iii) A colour camera} \revF{(Player One Neptune~664C, $2.9\,\mu$m pixels) on a 0.20~m Celestron RASA~8 astrograph (focal length 400~mm, f/2.0)}
\revY{recorded continuously at 100~frames~s$^{-1}$
(10~ms exposure, }\revF{$1.5\arcsec$}\revY{~pixel$^{-1}$); of its data only the clip of
flash I was available for this analysis.} We reduced these data with the same pipeline (aperture
radius scaled to the plate scale) and calibrated \revY{the 3.06~ms camera}
independently on the same standard star HD~117544, which is present in
\revY{its} field. \revY{To compare the photometry with the 16.67~ms
frames of Funabashi we used two quantities. The \emph{peak magnitude in
the 16.67~ms window} is obtained by summing the background-subtracted
flux of five consecutive 3.06~ms frames (15.3~ms, rescaled to 16.67~ms),
sliding this window over the light curve, and taking the maximum; it is
the quantity measured in the Funabashi peak frame when the flash happens
to be centred in that frame. The \emph{time-integrated magnitude} is
the magnitude of the total background-subtracted flux summed over all
frames of the event, expressed as if it had been received in a single
16.67~ms frame (i.e. $R_{\rm int}=\mathrm{ZP}_{16.67}-2.5\log_{10}
(\sum_{i}F_{i}\,t_{\rm exp}/16.67~\mathrm{ms})-kX$, with the zero point
$\mathrm{ZP}_{16.67}$ referred to a 16.67~ms exposure); it measures the
total luminous energy and is independent of the exposure time and of the
phase of the frames.} For the four flashes detected in both data sets (A, B, D, H) the
peak times agree to within $\pm12$~ms, and the Hiratsuka peak magnitudes
in the 16.67~ms window agree with the Funabashi values
with a mean difference of $-0.2$~mag and a scatter of 0.4~mag, while the
time-integrated magnitudes agree within 0.1--0.6~mag
(Fig.~\ref{fig:fujii}); the two stations therefore provide an
independent check of the zero point at the 0.2~mag level. The
instantaneous 3~ms peak magnitudes are 1.2--1.9~mag brighter than the
16.67~ms frame averages, showing that the flashes reach their maximum
and fade by a factor of several within the first few milliseconds, as
also seen in the 10~ms light curve of flash I, which decays over seven
consecutive frames (70~ms)\revY{; because no standard star is in the
field of the 100~fps camera, its zero point was fixed by requiring the
time-integrated magnitude of flash I to equal the Funabashi value, so
that only the shape and the peak of its light curve are independent}. \revV{Flash I is not detected in the 3.06~ms frames
at its predicted position, an unsaturated earthshine region
(Fig.~\ref{fig:multicam}c); the orientation of that camera was solved
from the four flashes it detected, to within $4.6\degr$.} The 3~ms single-frame detection limit of the
Hiratsuka camera ($R\simeq7.6$--8.4) explains the non-detection of the
fainter flashes C, E, and F ($R\ge9.1$), and one additional candidate was
recorded only at Hiratsuka (19:58:20~UT, $R\simeq8.5$ at 16.67~ms
equivalent, marginal; F$'$ in Fig.~\ref{fig:app2025}) at a time when
the Funabashi recording was interrupted. A comparison of the light-curve
shape with the impact terrain is given in Appendix~\ref{app:terrain}.

\revR{\textit{Observing-site conditions.}\label{sec:site}
The atmospheric extinction measured during the 2025 campaign
($k=0.2822$) is typical of a low-elevation site. The scintillation was
characterised with the formula of \citet{Young1967},
$\sigma_{\mathrm{scint}}=C_{Y}D^{-2/3}X^{7/4}e^{-h/8000\,\mathrm{m}}(2t)^{-1/2}$
(relative flux rms, with $D$ in cm and $t$ in s): the frame-to-frame
scatter of the HD~117544 photometry (17--19\% per 16.67~ms frame at
$X=2.0$--2.4) gives $C_{Y}=0.09$ (0.08--0.11; dimensionless), also
typical of such sites.} \revV{The image size, measured by Gaussian fits,
was $2.5\arcsec$--$3.5\arcsec$ (FWHM) for the unsaturated 2018 flashes
and about $5\arcsec$--$6\arcsec$ in 2025, when the Moon was observed at
airmass 2.0--2.5 ($5.7\arcsec$--$6.2\arcsec$ for the standard star in
single frames and $4.1\arcsec$--$5.5\arcsec$ for the four brightest
flashes); the $7.2\arcsec$ aperture radius contains the flash image in
both campaigns.}

\section{Peak images and light curves of all flashes}
\label{app:flashes}

\begin{figure*}
\centering
\includegraphics[width=\textwidth]{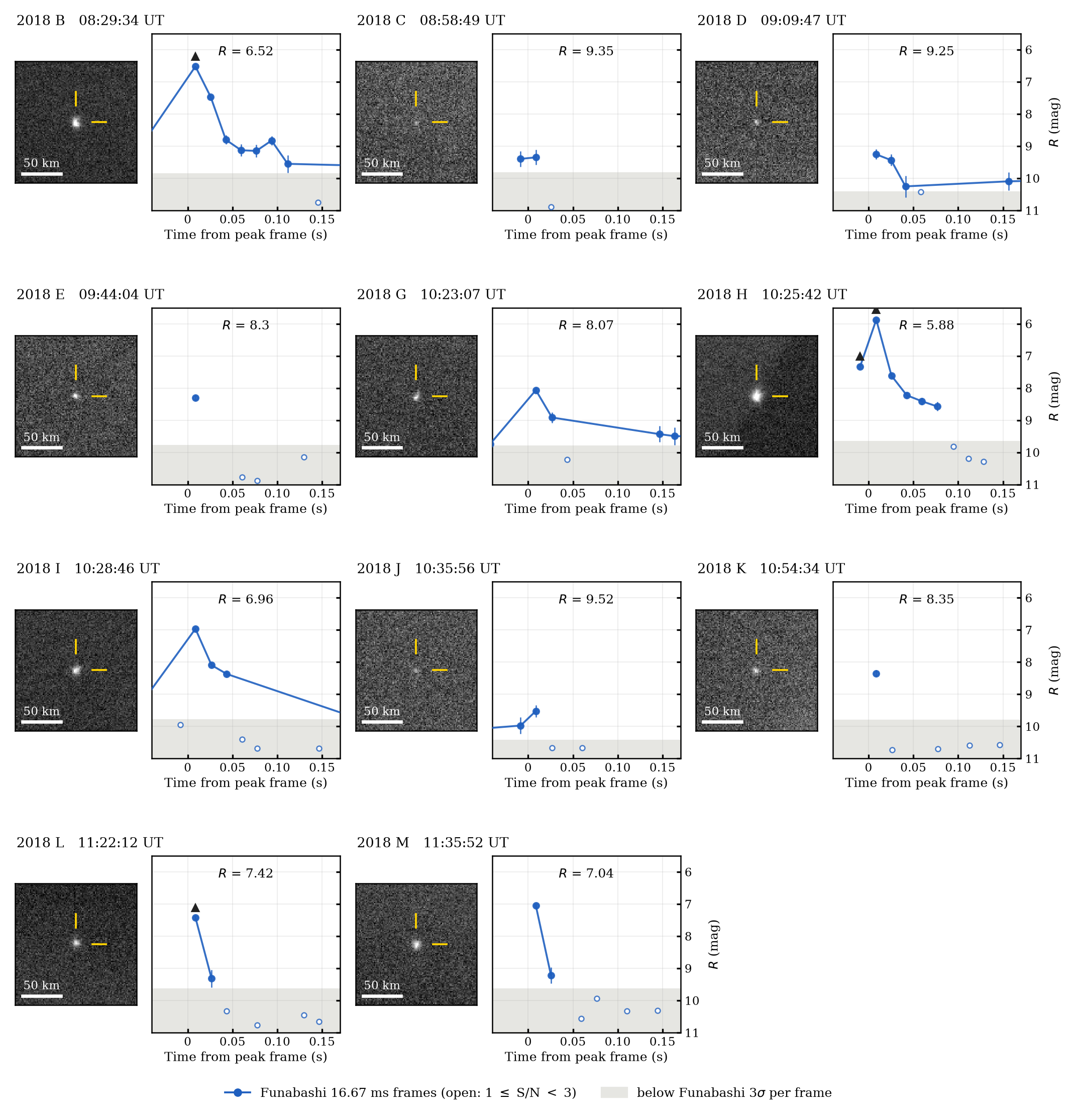}
\caption{Peak frames (left of each pair; $160\times160$~km, ticks point
to the flash) and calibrated $R$-band light curves (right) of the eleven
2018 flashes recorded at Funabashi. Points are the magnitudes of the
individual 16.67~ms frames at their mid-times, connected by lines (open
circles: $1\le$~S/N~$<3$); grey band: below the single-frame
$3\sigma$ limit; triangles: frames saturated in the 8-bit camera, whose
flux was recovered from the Gaussian wings. Times are in UT.}
\label{fig:app2018}
\end{figure*}

\begin{figure*}
\centering
\includegraphics[width=\textwidth]{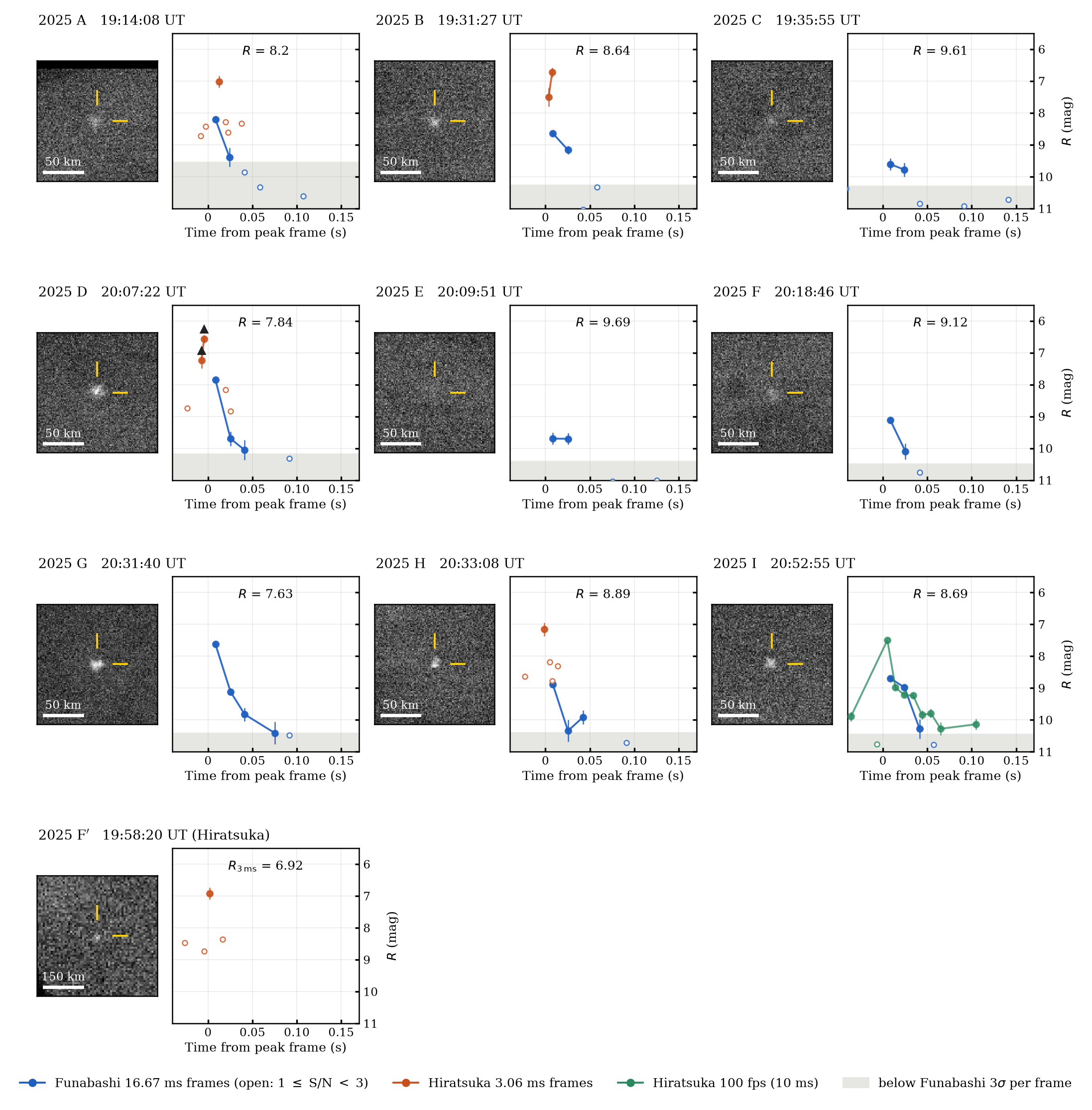}
\caption{As Fig.~\ref{fig:app2018} for the nine 2025 flashes recorded at
Funabashi, with the simultaneous Hiratsuka data (orange: 3.06~ms
frames; green: 100~fps camera, clock offset corrected). The last panel
shows the marginal candidate F$'$ seen only at Hiratsuka
($480\times480$~km cut-out of the 3.06~ms frame).}
\label{fig:app2025}
\end{figure*}

Figures~\ref{fig:app2018} and \ref{fig:app2025} show, for each flash,
the peak frame recorded with the 0.40~m GUNDAM telescope at Funabashi
together with its calibrated light curve. The images are the frames of
maximum light (2018: 8-bit SER frames; 2025: dark- and flat-corrected
frames), trimmed to $160\times160$~km on the Moon around the flash and
displayed in the native orientation of the camera with an arcsinh
stretch. The scale bars follow from the plate scale
($0.795\arcsec$~pixel$^{-1}$) and the topocentric distance of each
flash. The light curves give the calibrated Cousins $R$ magnitude of
each frame. The marginal 2025 candidate F$'$,
recorded only at Hiratsuka, is shown with the 3.06~ms Hiratsuka frame,
whose plate scale was derived from the radius of the lunar disc.

\section{Light-curve decay and target terrain}
\label{app:terrain}

\begin{figure*}
\centering
\includegraphics[width=0.9\textwidth]{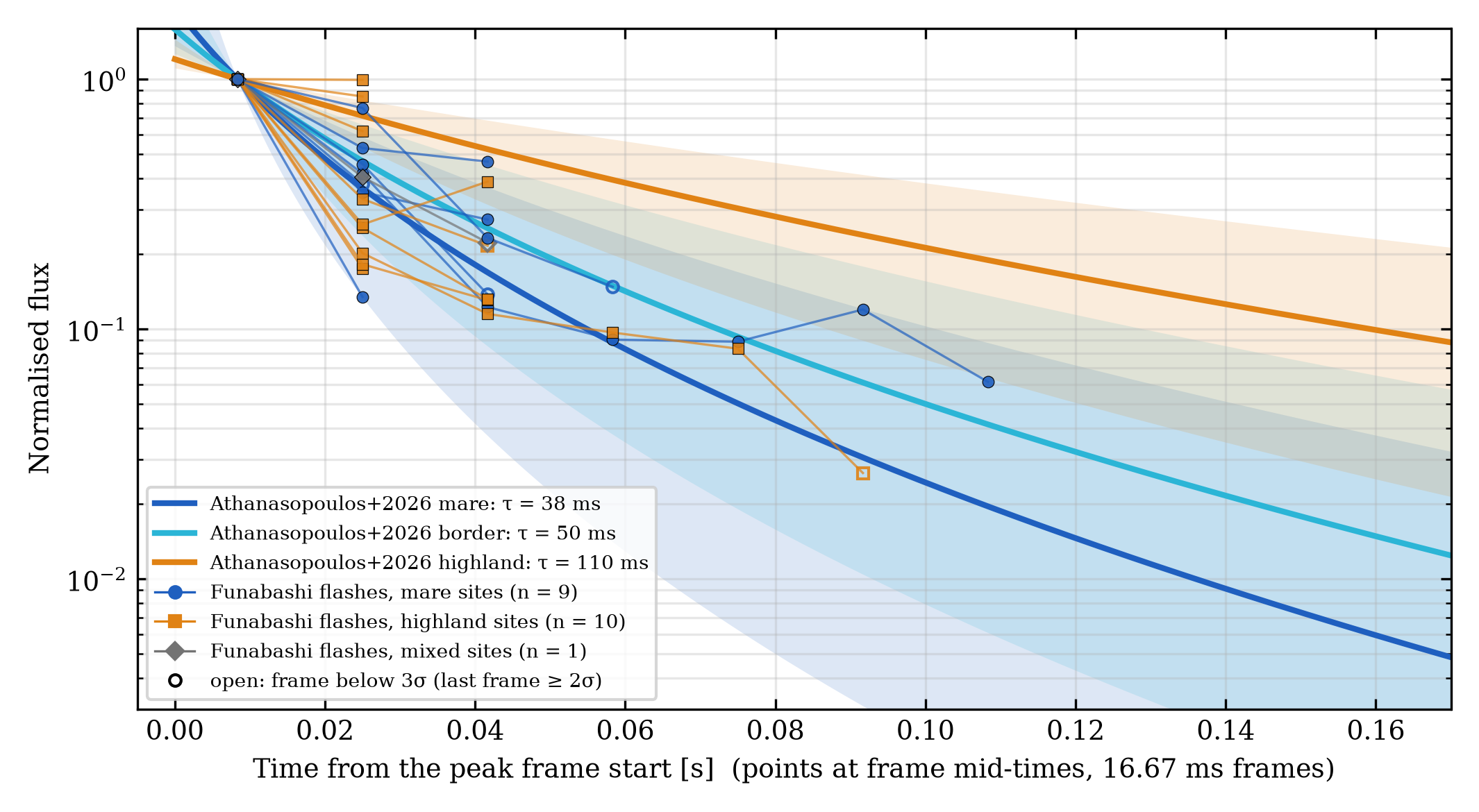}
\caption{Decay of the twenty Geminid flashes against the mare/highland
dichotomy of \citet{Athanasopoulos2026}. Each Funabashi light curve
(16.67~ms frames, points at the frame mid-times) is normalised to its
peak frame and coloured by the LROC WAC terrain class of the impact site
(blue circles: mare; orange squares: highland; grey diamonds: mixed);
open symbols are frames between $2\sigma$ and $3\sigma$. The thick lines
and bands are the single-size cooling model of
\citet{Athanasopoulos2026}, $F\propto[\exp(hc/\lambda kT)-1]^{-1}$ with
$T=T_{0}(1+(t-t_{\mathrm{ref}})/\tau)^{-1/3}$, $T_{0}=3100$~K,
$\lambda=700$~nm, $t_{\mathrm{ref}}=18$~ms, for their mare, border, and
highland values of $\tau$ with the quoted ranges, normalised to the
peak-frame mid-time.}
\label{fig:terrain}
\end{figure*}

The
exponential decay times of the twelve flashes that lasted at least three
frames are $\tau=13$--54~ms and correlate with the peak magnitude rather
than with the target terrain: a comparison of the nine mare and \revP{ten}
highland sites (\revP{one site is} mixed; classified from the LROC WAC albedo
at the re-measured coordinates) shows no significant difference in
duration, decay time, or in the ratio of the peak to the time-integrated
magnitude (Mann--Whitney $p=\revP{0.4}$--0.8).
Figure~\ref{fig:terrain} compares the normalised decay of all twenty
flashes with the single-size cooling model fitted by
\citet{Athanasopoulos2026} to NELIOTA flashes on mare ($\tau=38$~ms),
border ($\tau=50$~ms), and highland ($\tau=110$~ms) terrain: the
Funabashi light curves of both terrain classes fall between the mare and
border curves within their quoted ranges, and the highland events decay
as fast as the mare events. The
differing decay behaviour of individual light curves (e.g. the brighter
flash D fades faster than flash G) presumably reflects the impact
geometry and the phase of the flash within the 16.7~ms exposure window;
a systematic dichotomy of LIF light-curve shapes between mare and
highland targets has recently been reported by
\citet{Athanasopoulos2026}, but our centimetre-class impactors excavate
only the uppermost metres of regolith, which cover both terrain types.

\section{Alternative prescriptions and systematics of the flux, mass, and crater-size distributions}
\label{app:crater}

\textit{Shower association.} An alternative prescription for the Geminid probability of Sect.~\ref{sec:assoc} that extrapolates the shower and sporadic
populations separately to the mass of each impactor, with population
indices $r=2.6$ (Geminids) and $r=3.0$ (sporadics, mean impact speed
20~km\,s$^{-1}$), gives lower values, $p=\revP{0.41}$--$0.66$ (2018) and
$\revP{0.34}$--$0.57$ (2025), because the steeper sporadic magnitude
distribution of faint meteors, extrapolated over four orders of magnitude
in mass, predicts a sporadic flux of 10-g meteoroids comparable to the
Geminid flux.

\textit{Crater scaling.} For comparison with Eq.~(\ref{eq:crater}) we also evaluated the
$\pi$-group scaling of \citet{Holsapple1993} for dry sand in the gravity
regime ($K_{1}=0.132$, $\mu=0.41$, $\nu=0.4$, $g=1.62$~m\,s$^{-2}$, with
the vertical velocity component $v\sin\revY{\theta}$ and a rim-to-rim diameter of
2.6 times the apparent crater radius), which gives rim diameters 13--18\%
smaller than Eq.~(\ref{eq:crater}) for all twenty events; the two
relations thus agree to within their intrinsic uncertainty and both
predict metre-class craters.

\textit{Sporadic background models.} For the comparison in Fig.~\ref{fig:cummass} the bolide law of \citet{Brown2002}, $N(>E)=3.7E^{-0.9}$~yr$^{-1}$
($E$ in kt TNT), was converted to mass with $v=20.3$~km\,s$^{-1}$ and
divided by the Earth's surface area at 100~km altitude and by the Earth's
gravitational focusing factor (1.3 at 20~km\,s$^{-1}$); the
\citet{Grun1985} model is a flat-plate flux at 1~au and needs no such
correction, and the focusing by the Moon ($\sim$1\,\%) is neglected.

\textit{Difference between the two campaigns.} The difference between the two campaigns in Fig.~\ref{fig:cummass} is
largely a bright-end shortfall. The 2018 run recorded five flashes
brighter than $R=7.6$ in an exposure of \revN{$2.6\times10^{7}$}~km$^{2}$\,h;
scaled to the 2025 exposure of $1.75\times10^{7}$~km$^{2}$\,h, the
expected number is \revN{3.4}, and the Poisson probability of observing none is
\revN{3.5}\,\%. This $\sim$2$\sigma$ deficit is suggestive but not significant,
because the threshold is set a posteriori by the brightest 2025 flash.
Because the mass flux is dominated by the few largest bodies --
the brightest 2018 flash (\revN{H, $\sim$320~g}) alone carries 34\,\% of the
2018 mass flux -- the factor of \revN{2.4} between the two mass fluxes is the
expected consequence of this shortfall, not evidence of a change in the
stream.

The whole mass axis of Fig.~\ref{fig:cummass} scales
with the luminous efficiency as $\eta^{-1}$ while the slope does not:
for $\eta=5\times10^{-4}$ all masses increase by a factor of 2.8 and for
$\eta=6\times10^{-3}$ they decrease by a factor of 4.3, which shifts the
Geminid-to-background ratios quoted above by factors of 2.5 and 1/3.6,
respectively, but leaves the mass index, the size indices, and the
stability between the two years unaffected.

\begin{figure*}
\centering
\includegraphics[width=0.74\textwidth]{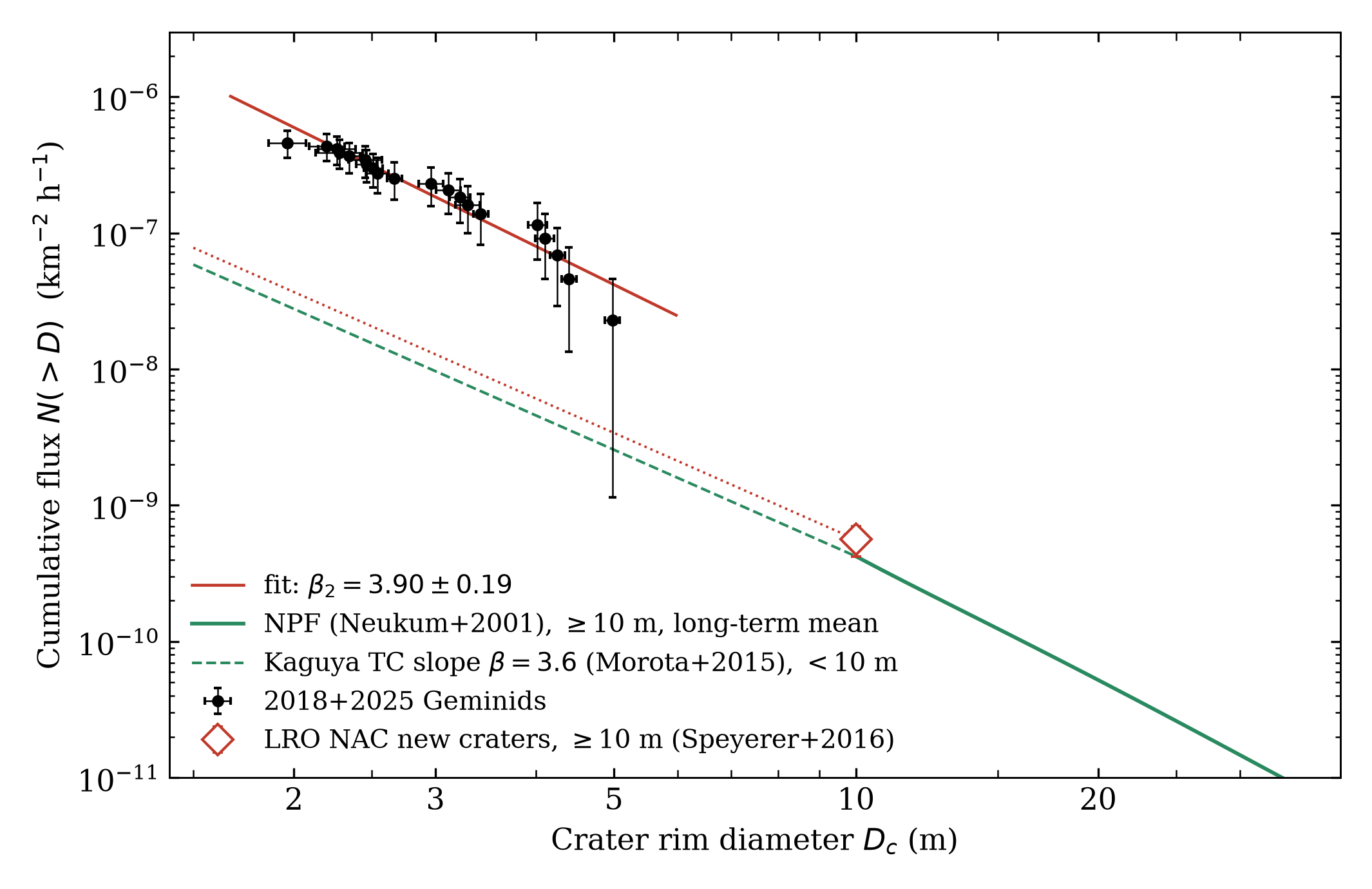}
\caption{As Fig.~\ref{fig:cumsizeDm}a for the predicted crater rim
diameter $D_{c}$ (\revY{Gault} scaling, Eq.~\ref{eq:crater}), with the
least-squares fit of Eq.~(\ref{eq:craterindex}) ($\beta_{2}=\revP{3.90}$).
Green lines: production function of \citet{Neukum2001} for $D\ge10$~m
(solid) and its extrapolation below 10~m with the slope $\beta=3.6$ of
\citet{Morota2015} (dashed). Red diamond: present-day rate of craters
$\ge10$~m from LRO NAC temporal image pairs \citep{Speyerer2016},
extrapolated below 10~m with the same slope (dotted).}
\label{fig:cumsizeDc}
\end{figure*}

\textit{Crater production rate.} In terms of craters
(Fig.~\ref{fig:cumsizeDc}), the production rate of 2--5~m craters during
the few hours of the Geminid maximum exceeds the long-term mean production
function extrapolated from the \citet{Neukum2001} function at 10~m with
the slope $\beta=3.6$ of \citet{Morota2015} by a factor of
$\sim$20; averaged over a year, the Geminid
contribution to the metre-scale cratering rate is therefore of order
half a per cent. Figure~\ref{fig:cumsizeDc} also shows the present-day production
rate of $\ge$10~m craters measured by \citet{Speyerer2016} from LRO NAC
temporal image pairs, which exceeds the \citet{Neukum2001} function by
33\,\%; \citet{Speyerer2016} attribute the excess mainly to secondary
impacts and to the limited statistics of their 222 new craters. A
half-per-cent annual contribution from the Geminids -- and at most a few
per cent from all annual showers combined -- cannot account for that
excess, so the contemporary cratering rate at metre scale is dominated by
the sporadic background and its secondaries, with the showers adding
short, intense bursts.

Two caveats apply to the crater-diameter distribution. First, the
crater-diameter index is not independent of the mass index: with
$D_{\rm c}\propto E_{\rm k}^{0.29}$ (Eq.~\ref{eq:crater}) and a fixed
impact velocity, $m\propto D_{\rm c}^{1/0.29}$, so that
$\beta_{2}-1=(s-1)/0.29$. For $s=1.76\pm0.17$ this predicts $\beta_{2}=3.6\pm0.6$,
between the least-squares (\revP{3.90}) and the maximum-likelihood
($3.3\pm0.5$) estimates; the differences reflect the
$(\sin\revY{\theta})^{1/3}$ term, the spread of distances, and the sensitivity
of the estimators to the faint end. The
comparison with the slope $\beta=3.6$ of \citet{Morota2015} therefore
tests the combination of our mass index and the crater-scaling exponent,
not an independent crater count. Second, the whole $D_{\rm c}$ axis
slides with the luminous efficiency as $\eta^{-0.29}$ and with the
choice of scaling law: for $\eta=5\times10^{-4}$ the crater diameters
increase by 35\,\%, whereas $\pi$-group scaling (see above)
reduces them by 13--18\,\%. Because the cumulative distribution is steep
($\beta_{2}-1\simeq2.3$--3.1), these shifts change the Geminid-to-background
ratio quoted above from $\sim$20 to between $\sim$10 and $\sim$50. The
conclusion that the shower produces a burst of metre-class craters one to
two orders of magnitude above the long-term mean rate is robust against
these systematics.

\section{Position of the LIF epochs in the stream and the ZHR extrapolation}
\label{app:stream}

Table~\ref{tab:offset} lists the position of the two LIF epochs relative
to the maxima of the millimetre-sized population and the fluxes
extrapolated from the visual ZHR (Sect.~\ref{sec:stable}). Because the extrapolation spans eight magnitudes, the predicted flux
scales as $r^{8.2}$, and the $\pm0.17$ uncertainty of the combined
index alone corresponds to a factor of $\sim$4 in either direction
(band in Fig.~\ref{fig:profile}b); the single-year indices
($s_{2018}=1.66\pm0.20$, $s_{2025}=2.10\pm0.37$) would spread the
extrapolation over more than an order of magnitude, so the comparison
is meaningful only with the combined index. The curves in Fig.~\ref{fig:profile}
are exponential models anchored to the published maxima, not the
measured GMN flux profile; the comparison rests only on the position of
the LIF windows relative to the millimetre-sized maxima and on the
post-maximum slope, both of which are well established.

\begin{table*}
\caption{Position of the two LIF epochs in the Geminid stream relative to the maxima of the millimetre-sized population, and the centimetre-sized flux measured there compared with the extrapolation of the visual (ZHR) flux.}
\label{tab:offset}
\centering
\footnotesize
\begin{tabular}{lccccc}
\toprule
\multicolumn{6}{l}{\itshape 2018: mid-window \revN{10:00} UT, $\lambda_{\sun,\mathrm{Moon}}=\revN{262.960}\degr$, ZHR at the LIF epoch \revN{51}} \\
\midrule
Reference maximum (mm-size) & $\lambda_{\max}$ (\degr) & $\Delta\lambda$ (\degr) & $\Delta t$ (h) & $\Delta x$ ($10^{6}$~km) & ZHR/ZHR$_{\max}$ \\
\midrule
IAU MDC established list \citep{Hajdukova2023} & 262.20 & $+0.76$ & $+18$ & 2.0 & 0.24 \\
GMN optical flux 2020--2021 \citep{Vida2022}   & 261.98 & $+0.98$ & $+23$ & 2.5 & 0.16 \\
IMO 2018 maximum \citep{Molau2020}             & 262.36 & \revN{$+0.60$} & $+14$ & \revN{1.5} & \revN{0.33} \\
\midrule
\multicolumn{6}{l}{\itshape Flux of $m\ge10$~g meteoroids at the 2018 LIF epoch (km$^{-2}$\,h$^{-1}$)} \\
Extrapolated from the ZHR with $r=2.6$ ($s=2.04$)                      & \multicolumn{5}{l}{$7.3\times10^{-8}$} \\
Extrapolated from the ZHR with $s=1.80$ (this work, least squares)      & \multicolumn{5}{l}{$4.3\times10^{-7}$} \\
Extrapolated from the ZHR with $s=1.76$ (this work, maximum likelihood) & \multicolumn{5}{l}{$6.0\times10^{-7}$} \\
Measured with LIFs (Table~\ref{tab:circumstances})                      & \multicolumn{5}{l}{\revN{$4.2\times10^{-7}$}} \\
Measured / extrapolated                                                 & \multicolumn{5}{l}{\revN{5.8} / 1.0 / 0.7} \\
\midrule
\multicolumn{6}{l}{\itshape 2025: mid-window 19:55 UT, $\lambda_{\sun,\mathrm{Moon}}=262.854\degr$, ZHR at the LIF epoch 42} \\
\midrule
Reference maximum (mm-size) & $\lambda_{\max}$ (\degr) & $\Delta\lambda$ (\degr) & $\Delta t$ (h) & $\Delta x$ ($10^{6}$~km) & ZHR/ZHR$_{\max}$ \\
\midrule
IAU MDC established list \citep{Hajdukova2023} & 262.20 & $+0.65$ & $+15$ & 1.7 & 0.30 \\
GMN optical flux 2020--2021 \citep{Vida2022}   & 261.98 & $+0.87$ & $+21$ & 2.2 & 0.20 \\
IMO 2018 maximum \citep{Molau2020}             & 262.36 & $+0.50$ & $+12$ & 1.3 & 0.40 \\
\midrule
\multicolumn{6}{l}{\itshape Flux of $m\ge10$~g meteoroids at the 2025 LIF epoch (km$^{-2}$\,h$^{-1}$)} \\
Extrapolated from the ZHR with $r=2.6$ ($s=2.04$)                      & \multicolumn{5}{l}{$6.1\times10^{-8}$} \\
Extrapolated from the ZHR with $s=1.80$ (this work, least squares)      & \multicolumn{5}{l}{$3.6\times10^{-7}$} \\
Extrapolated from the ZHR with $s=1.76$ (this work, maximum likelihood) & \multicolumn{5}{l}{$5.0\times10^{-7}$} \\
Measured with LIFs (Table~\ref{tab:circumstances})                      & \multicolumn{5}{l}{\revP{$5.2\times10^{-7}$}} \\
Measured / extrapolated                                                 & \multicolumn{5}{l}{8.4 / 1.4 / 1.0} \\
\bottomrule
\end{tabular}
\tablefoot{$\lambda_{\sun,\mathrm{Moon}}$ is the solar longitude (J2000) at
the middle of the observing window corrected for the position of the Moon
along the Earth's orbital motion ($-3.7$~h in 2018, $+3.1$~h in 2025; the
Moon was displaced perpendicular to the Earth's path by 31 and
$219\times10^{3}$~km). $\Delta t$ and $\Delta x$ are the
offsets in time and in distance along the orbit ($1\degr$ of solar
longitude $=1.0$~d $=2.57\times10^{6}$~km at the Earth's orbital
speed of 30.3~km\,s$^{-1}$). ZHR/ZHR$_{\max}$ is the fraction of the
visual peak rate at the LIF epoch for the post-maximum slope $B=0.81$.
The flux extrapolations use the ZHR at the LIF epoch, the
+6.5~mag meteoroid mass of \citet{Verniani1973}, the effective
collecting area of a visual observer for bright meteors
($2.4\times10^{5}$~km$^{2}$, zenith distances $<70\degr$), and the
gravitational-focusing correction between the Earth and the Moon; the
three ratios in the last row of each block refer to the three
extrapolations in order.}
\end{table*}

\section{Fragment size distribution and the Bennu analogue}
\label{app:frag}

Three remarks qualify the comparison of the size index of the Geminid pebbles
with the boulder populations of rubble-pile asteroids (Sect.~\ref{sec:large}). (i) As for the crater-diameter
index, $\beta_{1}$ is not an independent measurement: with
$D_{m}\propto m^{1/3}$, $\beta_{1}-1=3(s-1)$ exactly for either
estimator, so Fig.~\ref{fig:cumsizeDm} is the size-space representation of
the mass index. Its horizontal error bars propagate only the
photometric uncertainty; the luminous efficiency shifts the whole axis
as $\eta^{-1/3}$, by $+41$\,\% for $\eta=5\times10^{-4}$ (shaded band
in Fig.~\ref{fig:cumsizeDm}) and $-39$\,\% for $\eta=6\times10^{-3}$,
without changing $\beta_{1}$. (ii) The comparison with the
boulder populations spans more than two orders of magnitude in size
(pebbles of 2--6~cm against boulders of $>$5--8~m) and therefore tests
the slope, not the level. Cumulative size indices of $-2.5$ to $-3$ are
characteristic of fragments produced by impact disruption in laboratory
experiments \citep{Fujiwara1989}, and the boulder SFDs of Itokawa, Ryugu
and Bennu record the same fragmentation on their parent bodies; a
pebble population with the same index behaves as a fragment population
rather than as size-sorted dust. (iii) Thermal fracturing provides a
natural scale for such fragments. The diurnal thermal skin depth
$d=\sqrt{\kappa P/\pi}$ for Phaethon's rotation period $P=3.604$~h
and a thermal diffusivity $\kappa\simeq1$--$5\times10^{-7}$~m$^{2}$\,s$^{-1}$
(from the measured thermal inertia
$\Gamma=600\pm200$~J\,m$^{-2}$\,K$^{-1}$\,s$^{-1/2}$, \citealt{Hanus2016},
with $\rho c\simeq1.1\times10^{6}$~J\,m$^{-3}$\,K$^{-1}$ for the surface
layer) is $d\simeq2$--5~cm, so the diurnal temperature cycle of several
hundred kelvin at perihelion stresses only the top few
centimetres of the surface. Thermal-fatigue fragmentation
\citep{Delbo2014,Molaro2020} is therefore expected to release fragments
of a few centimetres, exactly the size range of the Geminid impactors
measured here (1.8--\revN{5.9}~cm).

The closest direct analogue of such ejecta is the particle-ejection
activity of the B-type near-Earth asteroid (101955) Bennu discovered by
OSIRIS-REx in 2019 \citep{Lauretta2019}. Bennu repeatedly released
millimetre- to centimetre-sized particles (diameters from $<$1\,cm up
to $\sim$10\,cm) at speeds of 0.07--3.3\,m\,s$^{-1}$, in events of up
to $\sim$200 particles that cluster in the late local afternoon
(15:00--18:00); the largest event (6 January 2019) carried
$\sim$0.5\,kg, and the total ejected mass is $\sim$10\,kg per 1.2-yr
orbit \citep{Lauretta2019,Hergenrother2020}. Of 313 tracked particles,
65\,\% were suborbital and 20\,\% completed at least one revolution,
while 15\,\% escaped directly on hyperbolic trajectories, Bennu's
escape speed being only $\sim$0.2\,m\,s$^{-1}$ \citep{Chesley2020}. Thermal fracturing driven by the large diurnal
temperature swings \citep{Molaro2020} and meteoroid impacts
\citep{Bottke2020} were identified as the most plausible drivers. The
Geminid impactors measured here ($D_{\rm m}=1.8$--\revN{5.9}\,cm, \revN{10--320}\,g)
fall squarely within the size range of the Bennu ejecta, demonstrating
that a B-type asteroid can shed centimetre-sized pebbles at the present
day without any volatile driver.

Two differences nevertheless matter
for Phaethon. First, with a diameter of 5.1\,km \citep{Hanus2016} Phaethon's escape
speed is $\sim$2--3\,m\,s$^{-1}$, an order of magnitude above Bennu's,
so Bennu-like launch speeds would leave most pebbles bound; escape
requires either the more violent thermal fracturing expected at
perihelion, where the surface exceeds 1000\,K and the diurnal stress
amplitudes far exceed those at Bennu ($q=0.90$\,au), or a faster
rotation in the past, close to the spin barrier
\citep{Jewitt2010,MacLennan2024}. Second, the mass budget: a Bennu-like
mass-loss rate of $\sim$10\,kg\,yr$^{-1}$ would deliver only
$\sim$10$^{5}$\,kg over the $\sim$10$^{4}$\,yr dynamical age of the
stream, whereas the Geminid stream mass is at least $1.6\times10^{13}$\,kg
\citep{Blaauw2017} and probably $10^{13}$--$10^{15}$\,kg
\citep{Ryabova2017}. The
centimetre-sized population we observe therefore cannot be a
steady-state, Bennu-like trickle; it must record one or a few episodes
of intense mass loss, consistent with the short ejection intervals
favoured by stream models \citep{Ryabova2012,Jo2024}, after which
thermal or rotational ejection at Phaethon's present-day perihelion has
been too weak to be detected beyond a faint, micron-dominated dust tail
\citep{Jewitt2010,Hui2017}. The Bennu observations thus establish the
mechanism and the particle sizes, while the LIF flux and the shallow
mass index set the magnitude of the past event(s).

The link between the LIF power law and the millimetre-sized population
(Fig.~\ref{fig:cumsizeDm}b) was evaluated as follows. Extrapolating the least-squares fit
(cumulative slope $-2.41$) from 1.9~cm down to the GMN limiting mass of
$2\times10^{-4}$~g ($D_{m}=0.5$~mm) predicts a flux of
$3.2\times10^{-3}$~km$^{-2}$\,h$^{-1}$, whereas the GMN optical flux at the Geminid maximum,
$2.0\times10^{-2}$~km$^{-2}$\,h$^{-1}$ \citep{Vida2022}, scaled to the
LIF epochs with the activity profile (16--20\,\% of the peak,
Table~\ref{tab:offset}) and corrected for the Earth's gravitational
focusing and for the perpendicular-area convention, is
$1.9$--$2.3\times10^{-3}$~km$^{-2}$\,h$^{-1}$ per unit lunar surface, a
factor of 1.4--1.7 lower; with the maximum-likelihood slope
($-2.3\pm0.5$) the same extrapolation gives
$1.7\times10^{-3}$~km$^{-2}$\,h$^{-1}$ and reproduces the GMN flux to
within 30\,\%.

\section{Registration of the video frames on the LROC mosaic}
\label{app:reg}

\revP{The position of the lunar disc and the orientation of the camera
(Sect.~\ref{sec:area}) were obtained as follows. The mosaic was projected into the frame for the topocentric
geometry at Funabashi (JPL DE421 ephemeris and lunar orientation,
perspective projection; disc radius of 1141 and 1125~pixels in 2018 and
2025) and cross-correlated with the median frame of the clip, solving for
the disc centre and the camera rotation angle; a circle of the ephemeris
radius fitted to the earthshine-lit limb provided the starting value of
the centre. The rotation angle is constant to $0.10\degr$
(rms) over the 2018 run and to $0.07\degr$ over 2025, as expected for a
camera fixed on an equatorial mount; the plate scale is recovered to
0.1\%, and the residual offsets between frame and mosaic, measured
independently in $\sim$30 sub-fields of each frame, are 0.4--1.1~pixels
rms (0.6--1.7~km at the disc centre; 3.7~pixels for the clip of 2018~L,
in which the field lies on the limb).}

\end{appendix}

\end{document}